\documentclass[preprint2]{aastex}

\usepackage{amsmath}
\usepackage{caption}
\usepackage{subcaption}
\usepackage{color}

\begin{document}


\title{Cluster dynamics largely shapes protoplanetary disc sizes}


\author{Kirsten Vincke and Susanne Pfalzner}
\affil{Max Planck Institute for Radio Astronomy, Auf dem H\"ugel 69, 53121 Bonn, Germany}
\email{kvincke@mpifr-bonn.mpg.de}



\begin{abstract}
  It is still on open question to what degree the cluster environment influences the sizes of protoplanetary discs surrounding young stars. Particularly so for the short-lived clusters typical for the solar neighbourhood in which the stellar density and therefore the influence of the cluster environment changes considerably over the first 10Myr. 
  In previous studies often the effect of the gas on the cluster dynamics has been neglected, this is remedied here. Using the code NBody6++ we study the stellar dynamics in different developmental phases - embedded, expulsion, expansion - including the gas and quantify the effect of fly-bys on the disc size.
  We concentrate on massive clusters $(M_{\text{cl}} \geq 10^3 - 6*10^4 M_{\text{sun}})$, which are representative for clusters like the ONC or NGC6611. We find that not only the stellar density but also the duration of the embedded phase matters.
  The densest clusters react fastest to the gas expulsion and drop quickly in density, here $98\%$ of relevant encounters happen before gas expulsion. By contrast, discs in sparser clusters are initially less affected but as they expand slower $13\%$ of discs are truncated after gas expulsion. For ONC-like clusters we find that usually discs larger than 500AU are affected by the environment, which corresponds to the observation that 200AU-sized discs are common. For NGC6611-like clusters disc sizes are cut-down on average to roughly 100AU.
  A testable hypothesis would be that the discs in the centre of NGC6611 should be on average $\approx$20AU and therefore considerably smaller than in the ONC.
\end{abstract}



 
\section{Introduction}
\label{sec:intro}

  Most stars are born in stellar clusters, which in turn form from dense cores in Giant Molecular Clouds (GMCs). At least for massive clusters ($M_{\text{cl}} > 10^{3}~M_{\sun}$), it is known that they are highly dynamical structures and follow well-defined evolutionary tracks, depending on their initial mass and size \citep{Pfalzner_Kaczmarek_2013b}.  At very young ages they are still embedded in their natal gas; the duration of this embedded phase is thought to last between $1-3~\text{Myr}$ for clusters in the solar neighbourhood \citep{Leisawitz_Bash_Thaddeus_1989, Lada_Lada_2003, Portegies_Zwart_McMillan_Gieles_2010}. 
  Comparing the gas and stellar content in nearby star forming regions, observations find that the fraction of gas in a GMC which is turned into stars (referred to as star formation efficiency - SFE) lies in the range of \mbox{$10\%-35\%$} \citep{Lada_Lada_2003}.
  Similarly, simulations that model the expansion history of massive clusters in the solar neighbourhood find that the SFE of these clusters must have been of the order $30\%$ \citep{Pfalzner_Kaczmarek_2013b}.
  In comparison, the SFE for an entire molecular clouds is much lower, only of the order of just a few per cent at most \citep[see e.g.][]{Murray_2011, Garcia_et_al_2014}.

  At the end of the star formation process the remaining gas is expelled through various mechanisms such as, for example, the explosion of a supernova \citep{Zwicky_1953, Pelupessy_Portegies_Zwart_2012}, bipolar stellar outflows \citep{Matzner_McKee_2000}, or stellar winds of the most massive stars \citep{Zwicky_1953, Dale_Ercolano_Bonnell_2012, Pelupessy_Portegies_Zwart_2012, Dale_Ercolano_Bonnell_2015}. 
  It is expected that ultimately supernovae will remove any remaining gas from the cluster, but probably other processes like wind are more important, as clusters are already found to be gas poor at 1-3 Myr, whereas even supernova with $25 M_{\text{Sun}}$ need already $7-8$ Myr until they explode.
  The gas expulsion itself is thought to happen on time scales smaller than or of the order of the dynamical times of the cluster \citep{Geyer_Burkert_2001, Melioli_de_Gouveia_dal_Pino_2006, Portegies_Zwart_McMillan_Gieles_2010}. Gas expulsion is supposed to happen earlier in massive than in low-mass clusters due to the larger number of high mass stars.

  The gas expulsion leaves the clusters in a supervirial state and they react by expanding with a simultaneous loss of a considerable portion of their members. The cluster dynamics after gas expulsion investigated thoroughly in the past \citep[e.g.][]{Lada_Margulis_Dearborn_1984, Goodwin_1997, Adams_2000, Geyer_Burkert_2001, Kroupa_Aarseth_Hurley_2001, Boily_Kroupa_2003a, Boily_Kroupa_2003b, Fellhauer_Kroupa_2005, Goodwin_Bastian_2006, Baumgardt_Kroupa_2007, Lueghausen_et_at_2012, Pfalzner_Steinhausen_Menten_2014, Pfalzner_Vincke_Xiang_2015}.

  Within the clusters their members interact with each other, influencing already formed protoplanetary discs. Processes like external photoevaporation \citep{Johnstone_Hollenbach_Bally_1998, Stoerzer_Hollenbach_1999, Scally_Clarke_2002, Matsuyama_Johnstone_Hartmann_2003, Johnstone_et_al_2004, Adams_et_al_2006, Alexander_Clarke_Pringle_2006, Ercolano_et_al_2008, Gorti_Hollenbach_2009, Drake_et_al_2009}, viscous torques \citep{Shu_Adams_Lizano_1987}, turbulent effects \citep{Klahr_Bodenheimer_2003}, and magnetic fields \citep{Balbus_Hawley_2002} are capable of reducing the discs in size, mass and/or angular momentum.

  However, here we concentrate on the effect of the gravitational forces acting during close stellar fly-bys shape the discs resulting in loss of angular momentum \citep[e.g.][]{Pfalzner_Olczak_2007} and/or mass \citep[e.g.][]{Clarke_Pringle_1993, Hall_1997, Scally_Clarke_2001, de_Juan_Ovelar_et_al_2012}. 

  Ideally one would simulate the entire cluster with each of the stars surrounded by a disc using smoothed particle hydrodynamics methods. In this case effects like viscous spreading of the discs and multiple fly-bys would all treated in a self-consistent way.
  Still, even with modern supercomputers this is extremely challenging. \cite{Rosotti_et_al_2014} performed a direct theoretical investigation of disc sizes in clusters by combining N-body simulations of a low-mass cluster (100 stars)  with smoothed particle hydrodynamics (SPH) simulations of protoplanetary discs and determined the disc sizes.
  Even for such a low-mass cluster he could only model the first $0.5~\text{Myr}$ of the development and had to make the artificial assumption of the stars to be of equal mass due to computational constraints. Thus, for the time being direct modelling of massive clusters and even more so a parameter study for those is completely out of the question.

  Therefore the standard procedure is a two step approach: First, \mbox{N-body} simulations of the cluster dynamics are performed where the fly-by history of each star is recorded and, second, results from parameter studies are used to post-process the data and determine the effect on the discs \citep[e.g.][]{Scally_Clarke_2001, Olczak_Pfalzner_Spurzem_2006, Pfalzner_Olczak_Eckart_2006, Olczak_Pfalzner_Eckart_2010,Steinhausen_Pfalzner_2014}. These studies concentrated on the disc frequency, average disc mass and angular momentum in the embedded phase of the cluster. 
  However, non of these studies considered the gas content as such or the effect that the gas expulsion process has on the cluster dynamics.
  Here we want to concentrate instead on the disc size, because (a) it is the most sensitive indicator for the cluster influence \citep{Rosotti_et_al_2014, Vincke_Breslau_Pfalzner_2015}, (b) with the advent of ALMA a direct comparison with observation is possible, and (c) it gives limits on the sizes of the potentially forming planetary systems that can be compared to exoplanetary systems.

  There have been a few studies that investigated the influence of fly-bys on the disc size. However, they were usually based on the results from parameter studies of fly-bys between equal-mass stars \citep{Kobayashi_Ida_2001, Clarke_Pringle_1993, Adams_2010}. A real cluster contains a wide spectrum of masses and therefore equal-mass fly-bys are the exception rather than the rule \citep{Pfalzner_Olczak_2007}. 
  Others proposed to convert the disc-mass criterion of \cite{Olczak_Pfalzner_Spurzem_2006} directly into a disc size \citep{de_Juan_Ovelar_et_al_2012}.
  Nevertheless,\cite{Breslau_et_al_2014} showed that this approach is error prone and devised a relation for the disc size after an fly-by which is valid over a large range of mass ratios between the star and the perturber.
 
  \cite{Vincke_Breslau_Pfalzner_2015}, in the following referred to as VBP15, used this more appropriate description of the effect of fly-bys on the disc size to perform study on embedded clusters of different mass and stellar density. They found that fly-bys in the embedded phase are capable of reducing discs to sizes well below $1\,000~\text{AU}$ and that the median disc size strongly depends on the stellar density.
  However, they again as all previous studies did not take into account the presence of the gas in the embedded phase and the effect of gas expulsion on the cluster dynamics.

  In contrast to previous studies, we include here the effect of the gas on the cluster dynamics and model all the evolutionary stages of the clusters self-consistently - the embedded phase, the gas expulsion, and the expansion phase. We quantify the differences between the fly-by history in the embedded phase and the expansion phase. More importantly, we will demonstrate how the differences in cluster dynamics and time scales influence the fly-by dynamics and the final disc-size distribution in dense and sparse clusters.

\section{Method}
\label{sec:method}

  \subsection{Cluster simulations}
  \label{sec:cluster_simulations}

    The cluster simulations are performed using the code Nbody6++ \citep{Aarseth_1973, Spurzem_1999, Aarseth_2003}. We model clusters of different mass, which is realised by performing simulations of clusters with different numbers of stars: \mbox{$1\,000$ (E0)}, \mbox{$2\,000$ (E1)}, \mbox{$4\,000$ (E2)}, \mbox{$8\,000$ (E3)}, \mbox{$16\,000$ (E4)}, and $32\,000$ (E52, E51). 
    However, the initial size of the clusters is kept fixed at a half-mass radius of $r_{\text{hm}} = 1.3~\text{pc}$ which allows to study clusters of different density. 
    Clusters depicted e.g. in \cite{Lada_Lada_2003} usually have somewhat smaller radii ($<1$~pc) as they still form stars. There are strong indications that clusters sizes increase with age during the star formation process and are typically about 1-2 pc by the time star formation is finished \citep{Kroupa_2005, Pfalzner_Kaczmarek_2013a, Pfalzner_et_al_2014}.
    
    Currently it is not clear to which extent massive clusters are subject to substructure. Any potentially existing substructure is quickly erased in the star formation phase \citep{Bonnell_Bate_Vine_2003, Parker_et_al_2014}, at the latest the gas expulsion will eliminate any left-over substructure in the presented extended clusters.
    For simplicity, we assume here an initial stellar number density distribution according to a relaxed, smooth King distribution \citep[][]{Olczak_Pfalzner_Eckart_2010} with a flat core which is representative for the Orion Nebula Cluster (ONC) which is just at the onset of gas expulsion and one of the best studied massive clusters in the solar neighbourhood.
    A detailed description of the density distribution, including an illustration of the initial density distribution as a function of the cluster radius (their Fig.~1), can be found in \cite{Olczak_Pfalzner_Eckart_2010}. 
    Any potentially existing substructure would make close encounters more common, so that the here presented results can be regarded as lower limits for the importance of the cluster environment on the protoplanetary discs.
    In contrast to VBP15 and most previous work here we take into account the potential of the gas component, too. The total mass of the system $M_{\text{cl}}$ is \mbox{$M_{\text{cl}} = M_{\text{stars}}+M_{\text{gas}}$} with $M_{\text{stars}}$ being the stellar component of the cluster, therefore, the gas mass is given by

    \begin{equation}
      M_{\text{gas}} = \frac{M_{\text{stars}} (1-\text{SFE})}{\text{SFE}}.
      \label{eq:gas_mass}
    \end{equation}

    where SFE is the star formation efficiency which is assumed to be $30\%$. Various studies have shown that such SFEs are characteristic for massive clusters like \mbox{NGC 2244}, \mbox{NGC 6611} etc. in the solar neighbourhood \citep[for example][]{Lada_Lada_2003}. The stars are initially still embedded in the remaining gas. Note that the gas density profile was chosen to be of Plummer form \citep{Steinhausen_phd_2013} with a half-mass radius similar to that of the stellar profile ($1.3~\text{pc}$), because King gas profiles lead to numerical difficulties.
  
    Apart from \cite{Rosotti_et_al_2014}, all previous studies of this kind did not include the gas component, including it here basically results in a different velocity dispersion compared to the gas-free case. It is assumed that the cluster is initially in virial equilibrium. The stellar velocities and the individual stellar masses are sampled randomly, the former from a Maxwellian distribution, the latter from the IMF by \cite{Kroupa_2002} with a lower stellar mass limit of $0.08~M_{\sun}$ and an upper mass limit of $150~M_{\sun}$. The embedded phase of clusters is thought to last between 1-3~Myr \citep{Leisawitz_Bash_Thaddeus_1989, Lada_Lada_2003, Portegies_Zwart_McMillan_Gieles_2010}. 
    Accordingly, we simulated clusters with an embedded phase lasting \mbox{$t_{\text{emb}}=2~\text{Myr}$}, but also performed an additional set of simulations for the densest cluster with \mbox{$t_{\text{emb}}=1~\text{Myr}$} (model E51). This allows us also to study how the length of the embedded phase influences the final distribution of protoplanetary disc sizes. 
    For a more detailed summary of the set-up parameters see \mbox{Table~\ref{tab:set-up_params}}.

    In contrast to previous work, we take into account that the gas expulsion process typically happens after $1-3~\text{Myr}$. The gas expulsion itself happens on short time scales, typically smaller than or of the order of several dynamical times $t_{\text{dyn}}$ of the cluster \citep{Geyer_Burkert_2001, Melioli_de_Gouveia_dal_Pino_2006, Portegies_Zwart_McMillan_Gieles_2010} which is given by

      \begin{equation}
	t_{\text{dyn}} = \left( \frac{GM_{\text{cl}}}{r_{\text{hm}}^3} \right)^{-1/2},
      \end{equation}

    The dynamical time scales for the cluster models E0-E52 are very short, between $0.8-0.14~\text{Myr}$, see Column~7 of Table~\ref{tab:set-up_params}.
    Therefore, and for better comparability of our cluster models, we assume the gas expulsion process in all clusters to be instantaneous. This immediate removal of the gas mass after \mbox{$t=t_{\text{emb}}$} leaves the cluster in a supervirial state, so that the cluster expands in order to regain virial equilibrium. We will discuss the consequences of such an instantaneous gas-expulsion on the results compared to a longer expulsion time scale in \mbox{Section~\ref{sec:discussion}}. We follow the cluster expansion until 10~Myr have passed since the cluster was fully formed. \\

    In each  simulation, the fly-by history for each individual star was tracked and the fly-by properties recorded. For each cluster model, a campaign of simulations with different random seeds was performed in order to improve statistics and minimise the effect of the initial individual set up of a cluster on the results. The number of simulations for each set-up is given in Column~3 of Table~\ref{tab:set-up_params}.

  \subsection{Disc size development}
  \label{sec:method_initial_disc_sizes}

    Ideally one would start out the simulation with an observed primordial disc size. However, observationally it is challenging to measure disc sizes directly especially in embedded clusters.
    In contrast to the disc fraction, disc size measurements are usually performed in (nearly) exposed clusters which have expelled most of their gas.
    For the best observed stellar cluster in the solar neighbourhood, the ONC, disc radii in the range from $\sim 27~\text{AU}$ up to $\sim 500~\text{AU}$ were found by several surveys \citep{McCaughrean_Odell_1996, Vicente_Alves_2005, Eisner_et_al_2008, Bally_et_al_2015}. However, the ONC is already 1 Myr old. Whether these measurements are representative for the primordial disc-size distribution or whether photoevaporation or fly-by processes have already altered the sizes remains unclear.
    In other clusters disc sizes up to several thousand AU have been reported. Therefore, there is no information about a typical initial disc size or a disc-size distribution in embedded stellar clusters.

    For this reason, and for simplicity, all discs in a cluster are set up with the same initial size $r_{\text{init}}$, ignoring any possible dependency of the disc size on the host mass \citep[cf.][]{Hillenbrand_et_al_1998, Vicente_Alves_2005, Eisner_et_al_2008, Vorobyov_2011, Vincke_Breslau_Pfalzner_2015}. 
    We performed a numerical experiment, setting the initial disc size to a very large value of \mbox{$r_{\text{init}} = 10\,000~\text{AU}$}. The interpretation of this large initial disc size will be discussed in \mbox{Section~\ref{sec:results}} and \mbox{Section~\ref{sec:discussion}}.

    In our simulations we determine the size of the protoplanetary discs around the cluster members after each stellar fly-by using the equation

    \begin{equation}
      r_{\text{disc}} = \begin{cases} 0.28 \cdot r_{\text{peri}} \cdot m_{12}^{-0.32}, & \mbox{if } r_{\text{disc}} < r_{\text{previous}} \\ 
				      r_{\text{previous}},                             & \mbox{if } r_{\text{disc}} \geq r_{\text{previous}}, 
		 \end{cases}
      \label{eq:breslau_disc_size}
    \end{equation}

    given by \cite{Breslau_et_al_2014}, where $m_{12}=m_{2}/m_{1}$ is the mass ratio between the disc-hosting star ($m_{1}$) and the perturber ($m_{2}$), $r_{\text{peri}}$ the periastron distance in AU, and $r_{\text{previous}}$ the disc size previous to the fly-by in AU. This equation is valid for coplanar, prograde, parabolic fly-bys. 
    This type of fly-by is more destructive than inclined, retrograde or hyperbolic fly-bys \citep[][]{Clarke_Pringle_1993, Heller_1995, Hall_1997, Pfalzner_et_al_2005, Bhandare_Pfalzner_2015}. However, the effect of inclined, retrograde and hyperbolic fly-bys is much less investigated. First results by \citeauthor{Bhandare_Pfalzner_2015} indicate that non-coplanar encounters have nevertheless a considerable effect on the disc size. Thus, the here presented result has to be regarded as lower limit of disc size, but will not be considerably smaller than it would be in the inclined case.

    Viscous forces, which might lead to disc spreading \citep{Rosotti_et_al_2014}, and self-gravity between the disc particles are neglected in this model because the discs are set up containing only mass-less tracer particles. 
    Every star in the cluster was surrounded by such a mass-less disc, therefore, each fly-by event is actually a disc-disc fly-by. Capturing of material from the disc of the passing star is disregarded in our approach as well. The formula above only holds for star-disc fly-bys, where only the primary hosts a disc. 
    Nevertheless, \cite{Pfalzner_Umbreit_Henning_2005} found that a generalisation of disc-disc fly-bys to star-disc fly-bys is valid as long as the discs have a low mass and not much mass is transferred between the two.
    For a more detailed description of the disc-size determination, its approximations, and the resulting influence on the results, see, \cite{Breslau_et_al_2014}. At the end of the diagnostic step the resulting fly-by and disc-size statistics are averaged over all simulations within one simulation campaign.

    Before presenting the results, we want to elucidate some definitions used in the following. We use the term ``fly-by'' in our study for gravitational interactions between two stars which a) reduce the disc size by at least $5\%$ ($r_{\text{disc}}/r_{\text{previous}} \leq 0.95$).
    The term ``strongest fly-by'' or ``disc-size defining fly-by'' describes the fly-by with the strongest influence on the disc in the whole simulation - or for certain periods of cluster evolution. Note that as Equation~\ref{eq:breslau_disc_size} takes into account the mass ratio of the perturber and the host star, the strongest fly-by is not necessarily the closest one.

\section{Results}
\label{sec:results}

\begin{figure}[t]
  \centering
  \includegraphics[width=0.48\textwidth]{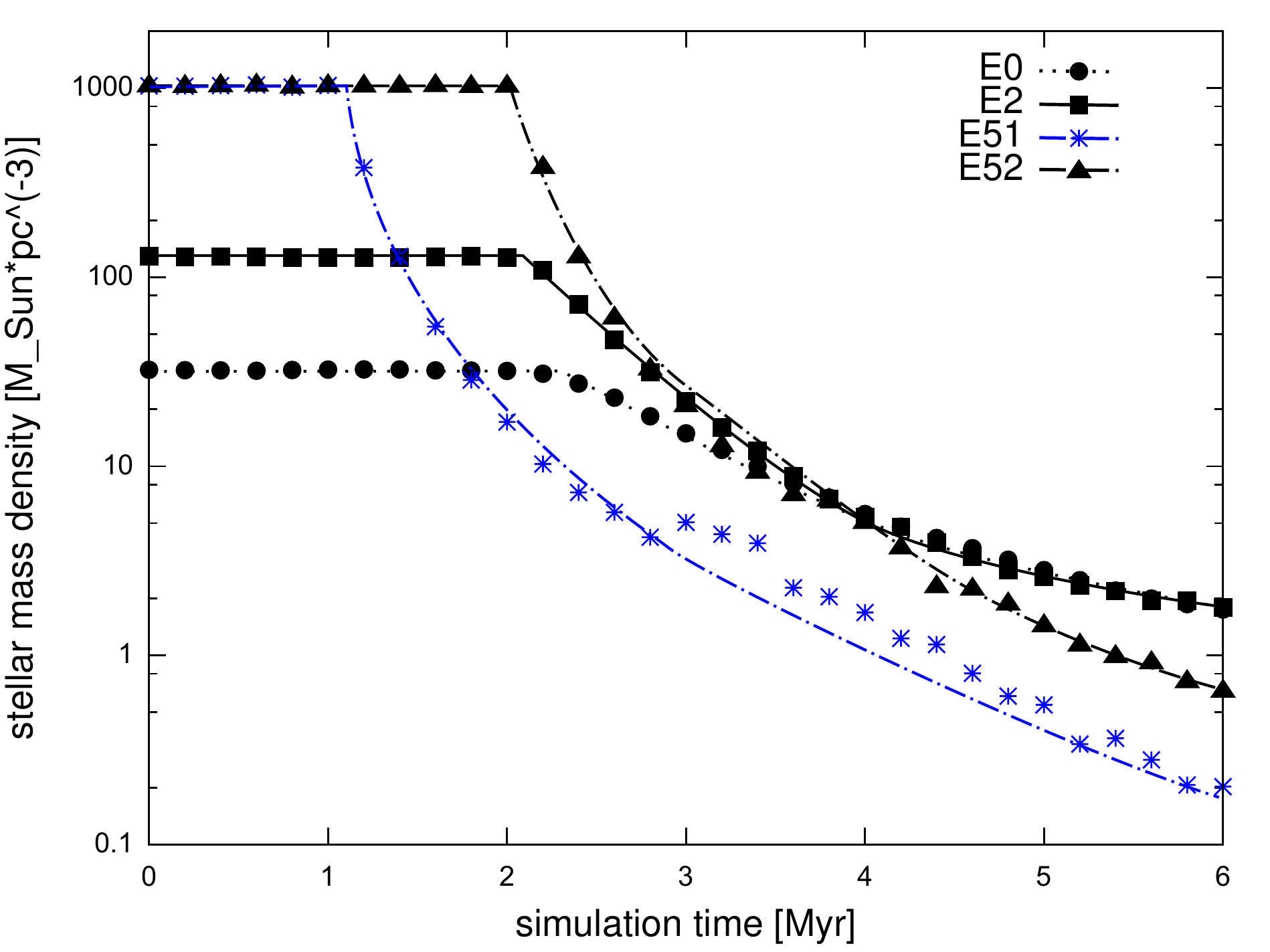}
  \caption{Stellar mass density within $1.3~\text{pc}$ (initial half-mass radius) as a function of time for clusters of different densities: \mbox{E0 (dots)}, \mbox{E2 (squares)}, \mbox{E51 (asterisks, blue)}, and \mbox{E52 (triangles)}. The duration of the embedded phase is \mbox{$t_{\text{emb}}=1~\text{Myr}$} for E51 and \mbox{$t_{\text{emb}}=2~\text{Myr}$} for all other models.}
  \label{fig:cluster_mass_density_vs_time}
\end{figure}

  The cluster evolution, namely the same mass and radius development - confirm previous work. However, here we have a closer look at the density evolution because this determines the here investigated fly-by history. Our simulations show that, as long as the clusters remain embedded in their natal gas (\mbox{$t_{\text{emb}}=2~\text{Myr}$}, for model E51 \mbox{$t_{\text{emb}}=1~\text{Myr}$}), the stellar mass density basically stays constant (see \mbox{Figure~\ref{fig:cluster_mass_density_vs_time}}). 
  When the gas is expelled instantaneously at \mbox{$t=t_{\text{emb}}$}, the clusters respond to the now supervirial state by expanding leading to a significant drop in the stellar density. 

  The more massive clusters regain their virial equilibrium much faster than the less massive clusters \citep{Parmentier_Baumgardt_2012, Pfalzner_Kaczmarek_2013a} due to their shorter dynamical time scales, see Table~\ref{tab:set-up_params}. As a result their stellar density declines faster than in the lower-mass clusters - the density in the most massive cluster (triangles and asterisks) drops to $10\%$ of its initial value already \mbox{$t=0.3~\text{Myr}$} after gas expulsion, whereas low mass clusters need up to \mbox{$t=2~\text{Myr}$} after gas expulsion for such a decline.

  Note that around \mbox{$t=3-4~\text{Myr}$} the cluster models E0, E2, and E52 are indistinguishable in terms of their stellar mass density within $1.3~\text{pc}$, while having a very different density history. \\

\begin{figure}[t]
  \centering
  \includegraphics[width=0.48\textwidth]{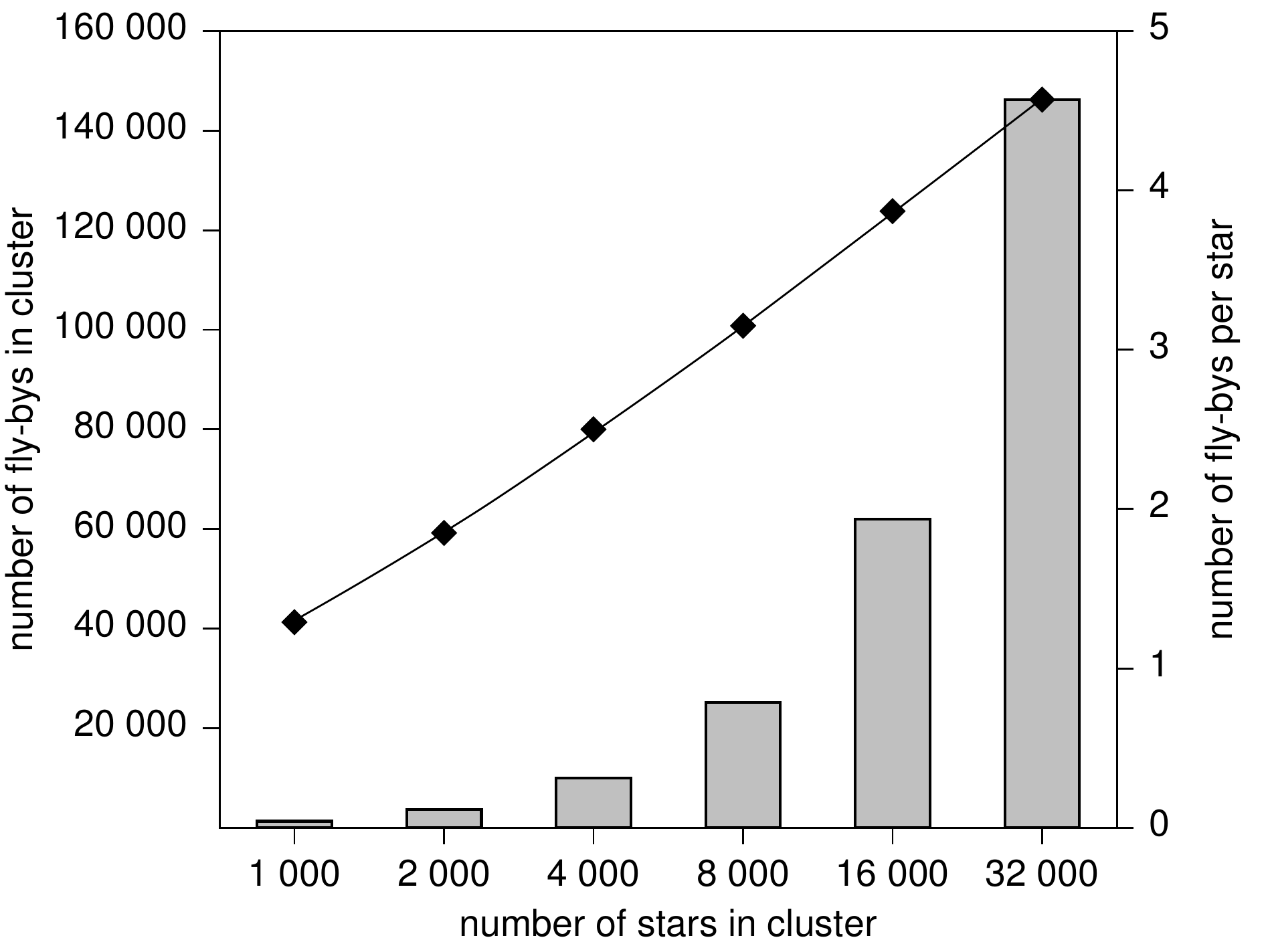}
  \caption{Number of fly-bys per cluster (grey boxes, left y-axis) and per star (black diamonds, right y-axis) for the different cluster models. The black line only serves to guide the eye.}
  \label{fig:dsc_enc_vs_cluster}
\end{figure}

  Naturally, the total number of fly-bys increases with cluster density which in our case is equivalent to the cluster mass. In the least dense cluster roughly $1\,300$ fly-bys that change the disc size take place during the $10~\text{Myr}$ simulated here whereas in the densest cluster model the number of fly-bys is approximately $150\,000$ (see \mbox{Figure~\ref{fig:dsc_enc_vs_cluster}}). 
  However, this increase is by far not as much as one would expect from a roughly 32 times higher density of models E51 and E52 in the embedded phase (see \mbox{Figure~\ref{fig:cluster_mass_density_vs_time}}). 
  The reason is that we only consider fly-bys that lead to a smaller disc size than previous to the fly-by, see Sect.~\ref{sec:method}. For the dense clusters the disc sizes are reduced very quickly to very small sizes so that even closer disc-size changing fly-bys are rare at later times.
  Similarly, the number of fly-bys per star increases with cluster density. In model E0 each star undergoes on average a little more than one disc-size changing fly-by as defined above, whereas in model E52 its between four and five. Although the difference in density (within the half-mass radius) between these two models is almost a factor of 100, the average number of fly-bys increases almost linearly by a factor of four due to the criterion mentioned above.

\begin{figure*}[t]
  \centering  
  \begin{subfigure}[t]{0.48\textwidth}
    \includegraphics[width=\textwidth]{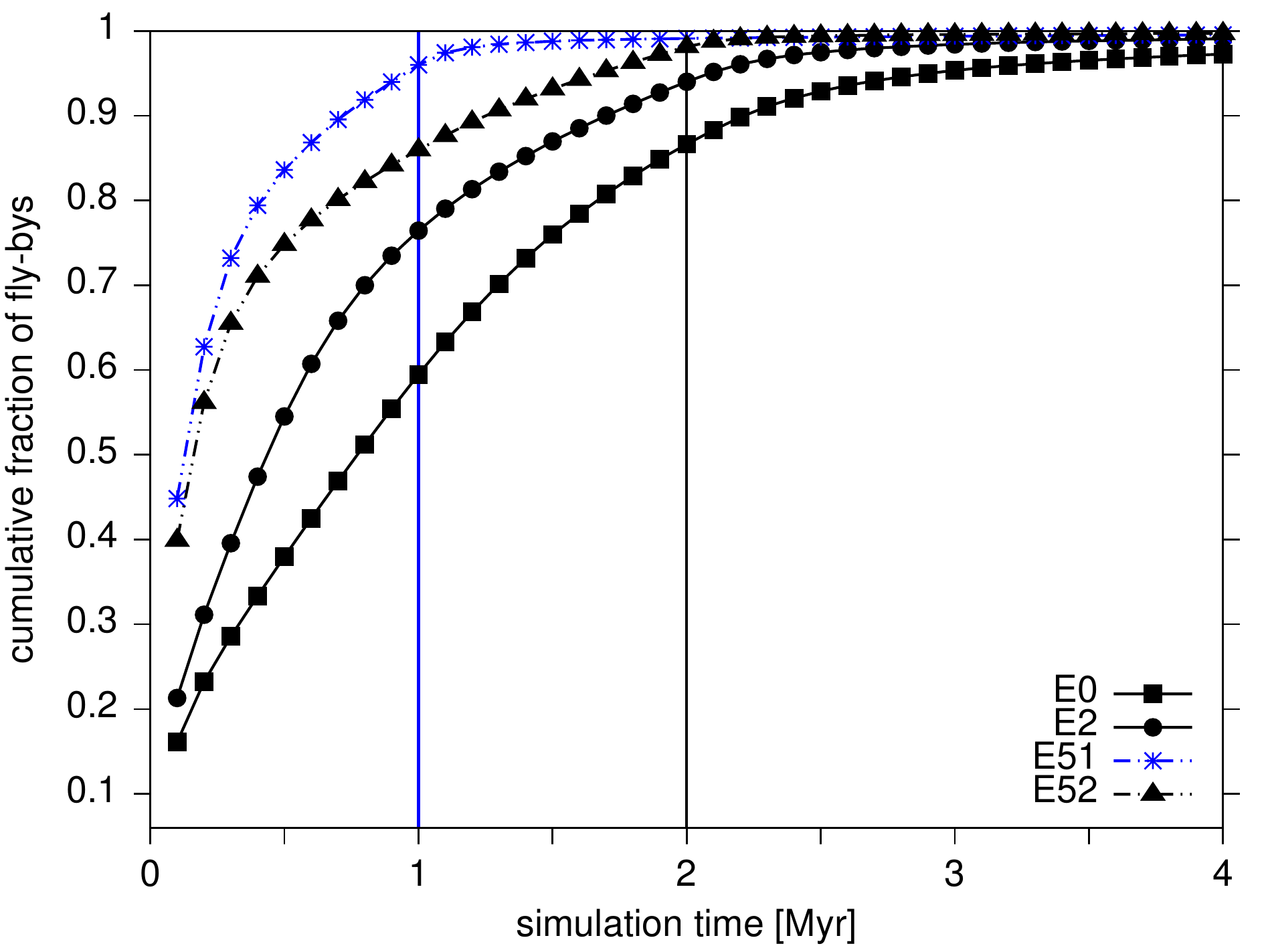}
  \end{subfigure}
  \hfill
  \begin{subfigure}[t]{0.48\textwidth}
    \includegraphics[width=\textwidth]{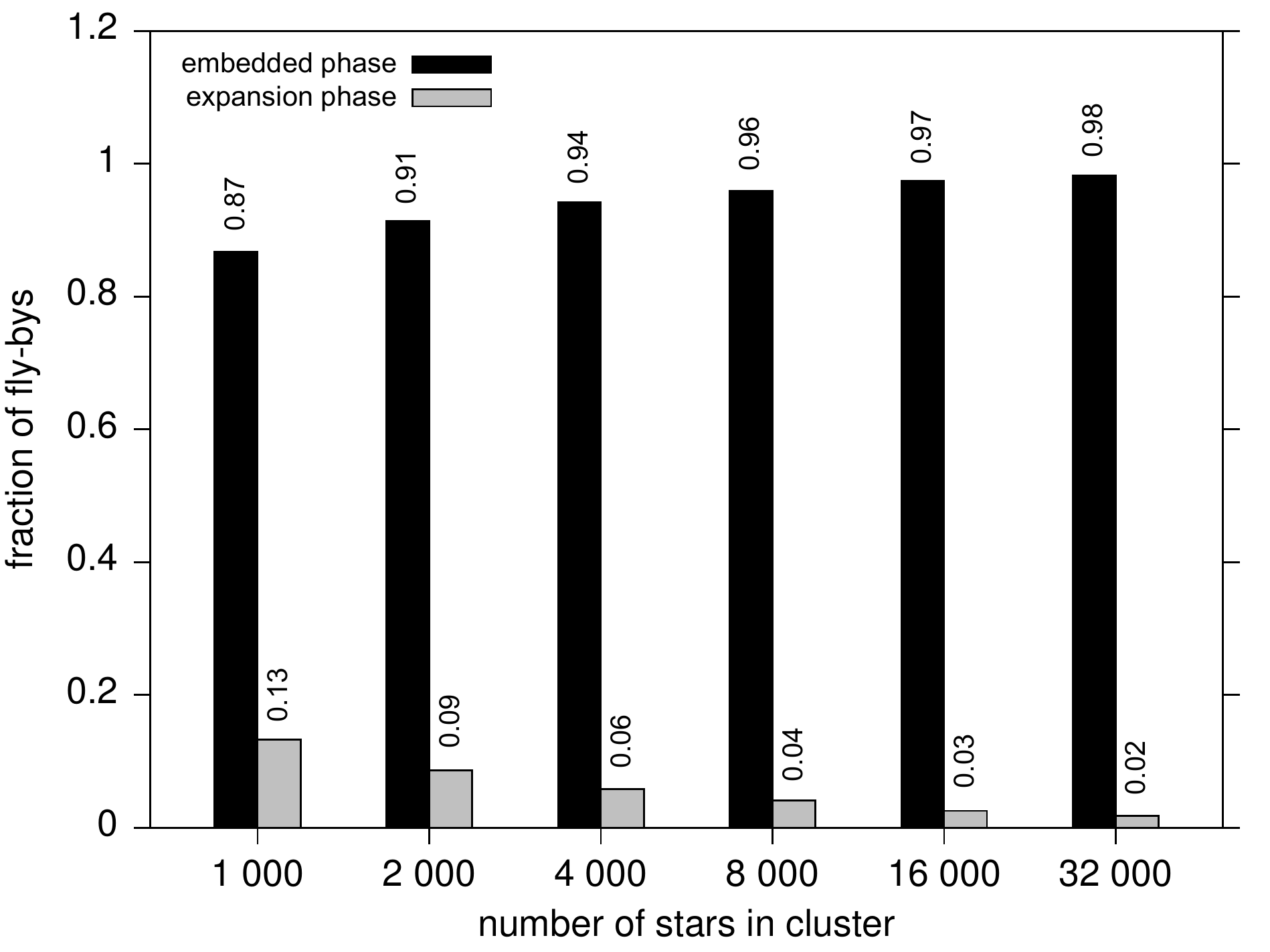}
  \end{subfigure}
  \caption{(a) Cumulative fraction of fly-bys as a function of time for the cluster models \mbox{E0 (squares)}, \mbox{E2 (dots)}, \mbox{E51 (asterisks, blue)}, and \mbox{E52 (triangles)}. The vertical lines depict the points in time of gas expulsion for model E51 ($1~\text{Myr}$, dotted blue) and for all other models ($2~\text{Myr}$, solid black). (b) Fraction of fly-bys as a function of the number of stars in the cluster for the embedded phase (black) and the expansion phase (grey).}
  \label{fig:enc_vs_time}
\end{figure*}

  This is also reflected in the temporal development of the number of disc-size changing fly-bys. \mbox{Figure~\ref{fig:enc_vs_time}~(a)} depicts the fly-by history in the different cluster models. It shows the cumulative fraction of fly-bys as a function of time, where the vertical lines mark the time of gas expulsion ($1~\text{Myr}$ for model E51, dotted blue, $2~\text{Myr}$ all other models, solid black). 
  The steeper slopes for the most massive clusters indicate that the discs are processed faster. For example, more than $50\%$ of all disc-size reducing fly-bys in \mbox{model E52} occur within the first \mbox{0.2~Myr} whereas in \mbox{model E0} it takes four times as long \mbox{($\sim 0.8~\text{Myr}$)} for the same portion of fly-bys to happen.
  
  As to be expected, the majority of fly-bys happens in the dense embedded phase. However, there are differences between the different cluster types, see \mbox{Figure~\ref{fig:enc_vs_time}~(b)}. Whereas in the most massive clusters disc-size changes happen nearly exclusively ($\sim 98\%$) in the embedded phase (black) - and most even within the first few 100,000 years - in the least dense clusters only $87\%$ of all disc size changes occur in this phase. 
  The reason is that in the latter case the density decreases more slowly so that a higher fraction of about one seventh of disc-size reducing fly-bys happen in the expansion phase (grey).

  Obviously the length of the embedded phase plays an important role. In \mbox{model E51}, the gas is expelled after $1~\text{Myr}$, whereas for \mbox{model E52} the gas expulsion happens after $2~\text{Myr}$. The earlier drop in cluster mass density in \mbox{model E51} results in the total number of fly-bys in \mbox{model E52} being roughly $15\%$ larger than in \mbox{model E51} ($151\,000$ compared to $131\,000$). 
  Again the reason why the number of fly-bys does not double for a twice as long embedded phase is that most of the discs have been reduced to a small disc size during the first Myr and therefore the cross section for a disc-size changing fly-by has been reduced. This means the early embedded phases largely determine the disc sizes. \\

\begin{figure}[t]
  \centering
  \includegraphics[width=0.48\textwidth]{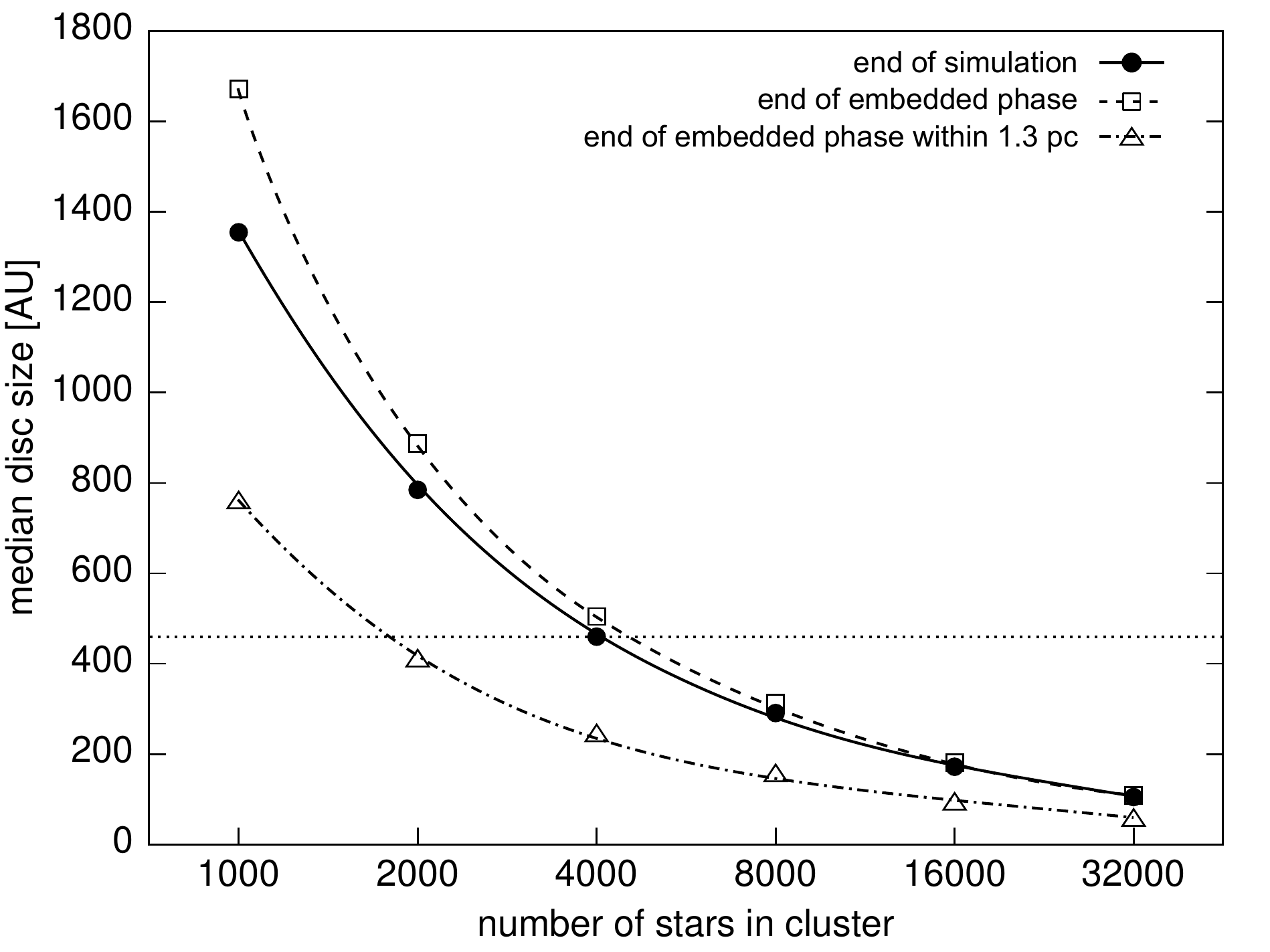}
  \caption{Overall median disc size for all stars for different cluster models at the end of the simulation ($10~\text{Myr}$, dots), at the end of the embedded phase ($2~\text{Myr}$, squares), and at the end of the embedded phase within a sphere of $1.3~\text{pc}$ (initial half-mass radius, triangles).}
  \label{fig:mean_disc_size_vs_cluster}
\end{figure}

  The distinct clusters have very different influence on their protoplanetary discs, reflected for example in the overall median disc size (\mbox{Figure~\ref{fig:mean_disc_size_vs_cluster}}).
  This median disc size is about thirteen times smaller in \mbox{model E52} ($32\,000$ stars) than in  \mbox{model E0} ($1\,000$ stars) as not only the number of fly-bys increases significantly with cluster density, but they are on average also closer or the mass ratio is higher.
  For the densest clusters most fly-bys happen at the beginning of the embedded phase, thus, the median disc sizes are nearly the same at the end of the embedded phase \mbox{($\sim 108~\text{AU}$, open squares)} and at the end of the simulations \mbox{($\sim 104~\text{AU}$, dots)}. However, for \mbox{model E0} the median disc size is significantly larger \mbox{($\sim 1\,670~\text{AU}$)} at the end of the embedded phase ($2~\text{Myr}$) than at the end of the simulations \mbox{($\sim 1\,350~\text{AU}$)} as roughly one seventh of the close fly-bys occur in the expansion phase. \\
  It is important to note that we do not expect real discs to be generally as large as nearly $1\,700~\text{AU}$. The median disc size here only reflects the degree of the environment's influence on the discs. For example, as long as the discs are initially $> 100~\text{AU}$, they are reduced in size in the densest cluster model. By contrast, in the ONC model only discs that are initially larger than $\sim 500~\text{AU}$ are affected. A real initial disc size distribution would be necessary to further constrain this, for a more detailed discussion see Sect.~\ref{sec:discussion}.

\begin{figure}[t]
  \centering
  \begin{subfigure}[t]{0.48\textwidth}
    \includegraphics[width=\textwidth]{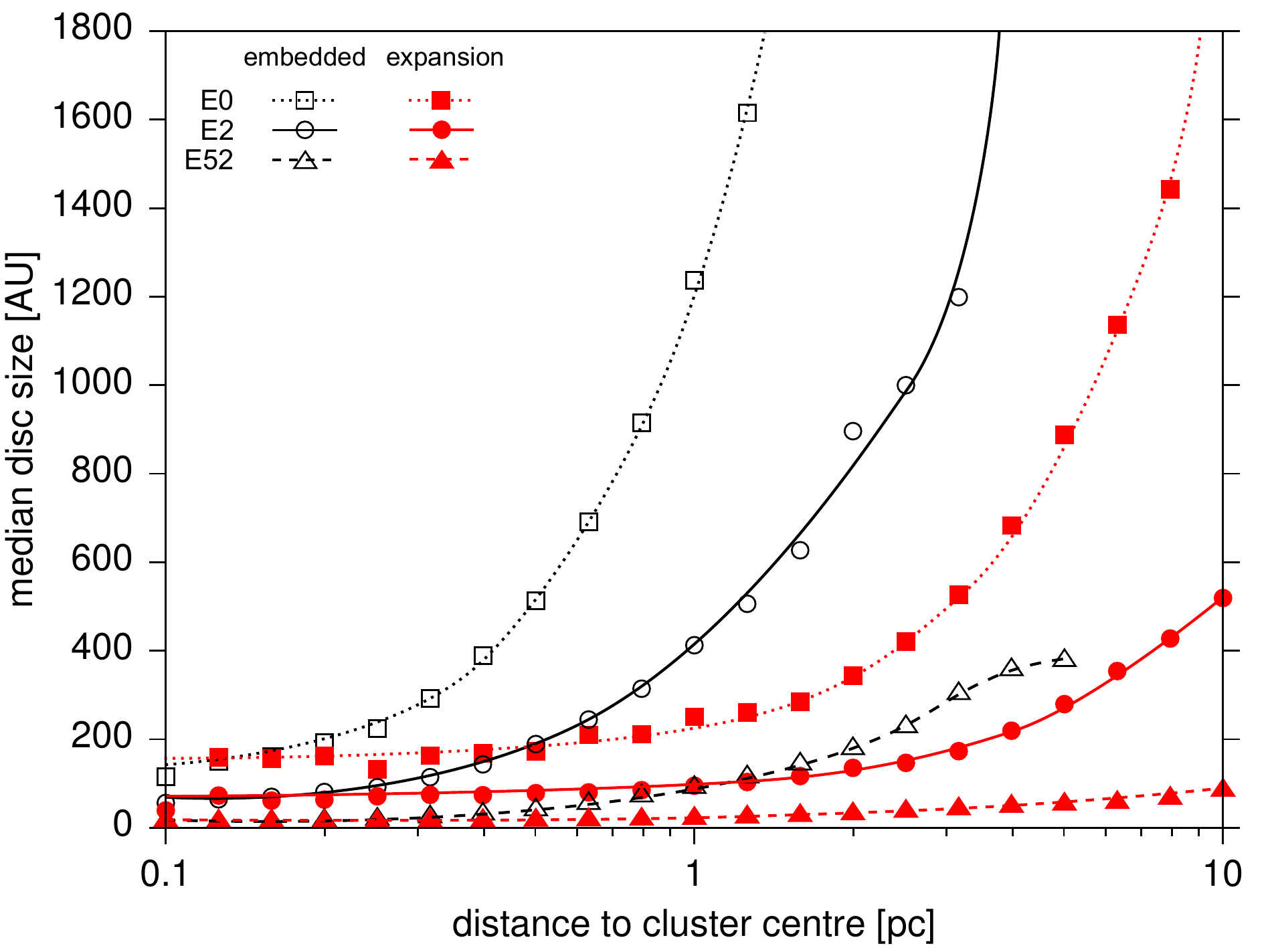}
  \end{subfigure}
  \hfill
  \begin{subfigure}[t]{0.48\textwidth}
    \includegraphics[width=\textwidth]{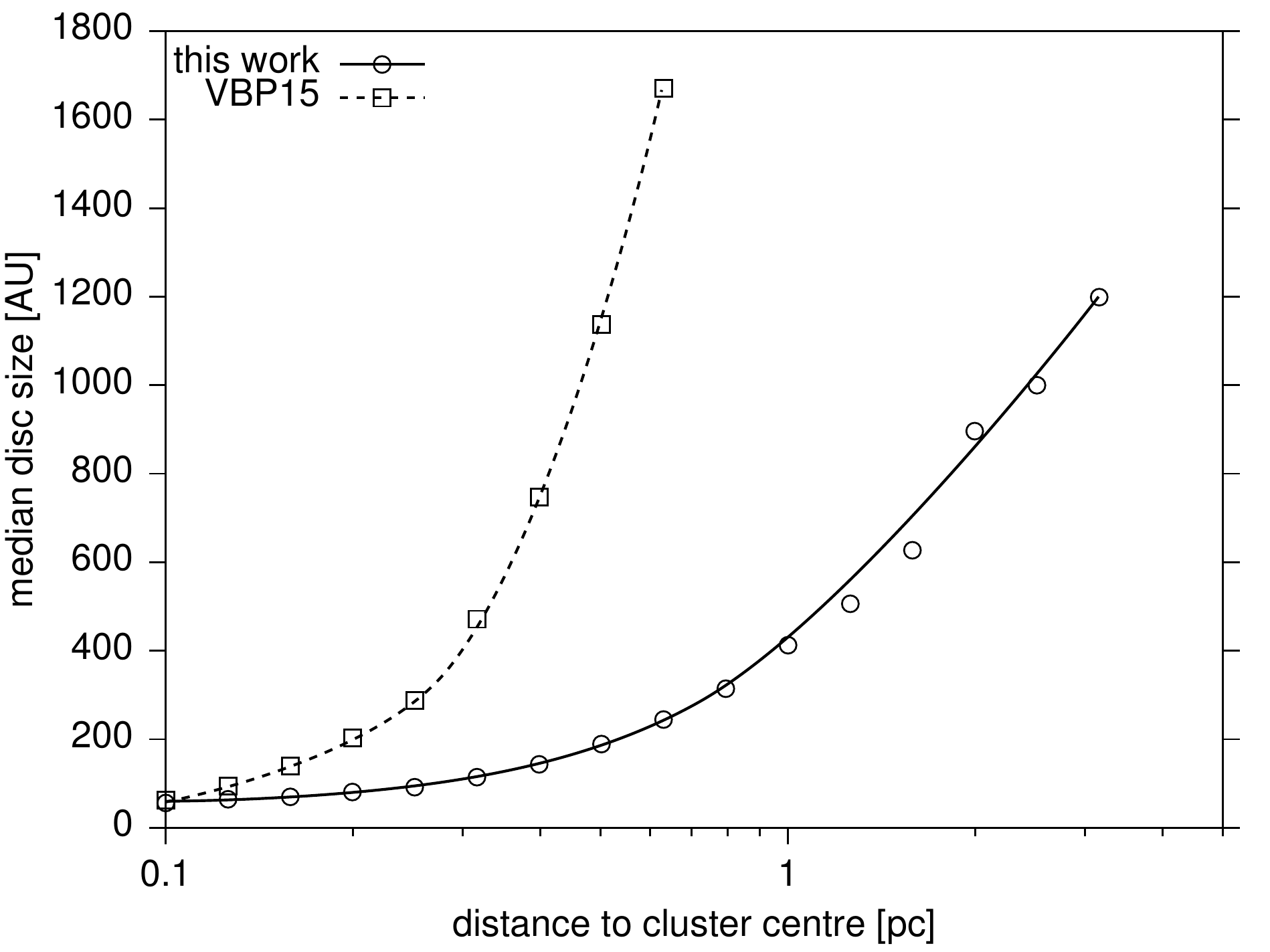}
  \end{subfigure}
  \caption{Median disc size as a function of the distance to the cluster centre (a) for different cluster models (E0 squares, E2 circles, and E52 triangles) at the end of the embedded phase ($2~\text{Myr}$, black symbols) and at the end of the simulation covering the embedded, the gas expulsion, and the expansion phase ($10~\text{Myr}$ (red symbols); 
  (b) for the ONC \mbox{model (E2)} in this work, i.e. with gas mass (circles, same as in (a)), and in VBP15, i.e. without gas mass (squares). The lines in (a) and (b) only serve to guide the eye.}
  \label{fig:mean_disc_size_vs_distance_to_cluster_center}
\end{figure}

  What does the spatial disc size distribution at the end of embedded phase look like?
  \mbox{Figure~\ref{fig:mean_disc_size_vs_distance_to_cluster_center}(a)} shows the median disc size as a function of the distance to the cluster centre of the stars for different cluster models at $t=2~\text{Myr}$ (open black symbols). In the inner part of the ONC-like cluster (E2) for example, within a sphere of the initial half-mass radius ($1.3~\text{pc}$) the median disc sizes are considerably smaller than for the clusters  outskirts. The difference is even larger when one compares the extremes - the median disc size rises from $50~\text{AU}$ at $0.1~\text{pc}$ to $2\,000~\text{AU}$ at $4~\text{pc}$. This is due to the higher density in the cluster core and the resulting higher fly-by frequency. 

  These trends have already been seen in simulations where the gas content was neglected (VBP15), however, there are quantitative differences.
  \mbox{Figure~\ref{fig:mean_disc_size_vs_distance_to_cluster_center}(b)} compares the median disc size as a function of the distance to the cluster centre after $2~\text{Myr}$ of simulation time for the ONC cluster model (E2) obtained in VBP15 (open squares) and in this work (circles).
  Including the gas mass explicitly leads to a higher velocity dispersion in the embedded phase and thus stronger encounters. Therefore, the median disc sizes presented in this work are much smaller than in VBP15. For example, at the rim of the cluster core (0.3~pc) the median disc size in the work here is more than a factor of four smaller than in VBP15 ($\sim 108~\text{AU}$ compared to $\sim 470~\text{AU}$). 
  At a distance of 1~pc the situation is even more extreme, as in our work the median disc size is roughly $400~\text{AU}$ whereas in VBP15 more than half of the discs are not influenced at all and still retain their initial size.
  
  If we consider the first $10~\text{Myr}$ of cluster evolution, which includes the embedded, gas-expulsion, and expansion phase (black symbols in \mbox{Figure~\ref{fig:mean_disc_size_vs_distance_to_cluster_center}}),
  In general, the denser the cluster is, the smaller the median disc size remains.
  Nonetheless, after $10~\text{Myr}$ the median disc size is nearly constant (at least for models E2 and E52) within $3~\text{pc}$ from the cluster centre. 
  This is not so much due to mixing, but basically mostly caused by the expansion of the cluster - the value of the median disc size in the cluster outskirts is now similar to that in the centre at the end of gas expulsion ($2~\text{Myr}$). 
  While during the embedded phase most stars do not move significantly in radial directions and the dependence of the median disc size on the distance to the cluster centre is preserved, after gas expulsion only about $10\%$ of stars remain bound to the cluster and the rest leaves the cluster very quickly \citep{Fall_Chandar_Whitmore_2009,Dukes_Krumholz_2012}. 
  The still bound stars largely move to positions more distant from the cluster centre than they were originally. That is the reason why the median disc size throughout the cluster in the expansion phase is similar to the median disc size in the inner cluster region shortly before gas expulsion. This means that when older clusters are observed they work like a microscope showing us the central area of enlarged versions of younger clusters. \\

\begin{figure}[t]
  \centering
  \includegraphics[width=0.48\textwidth]{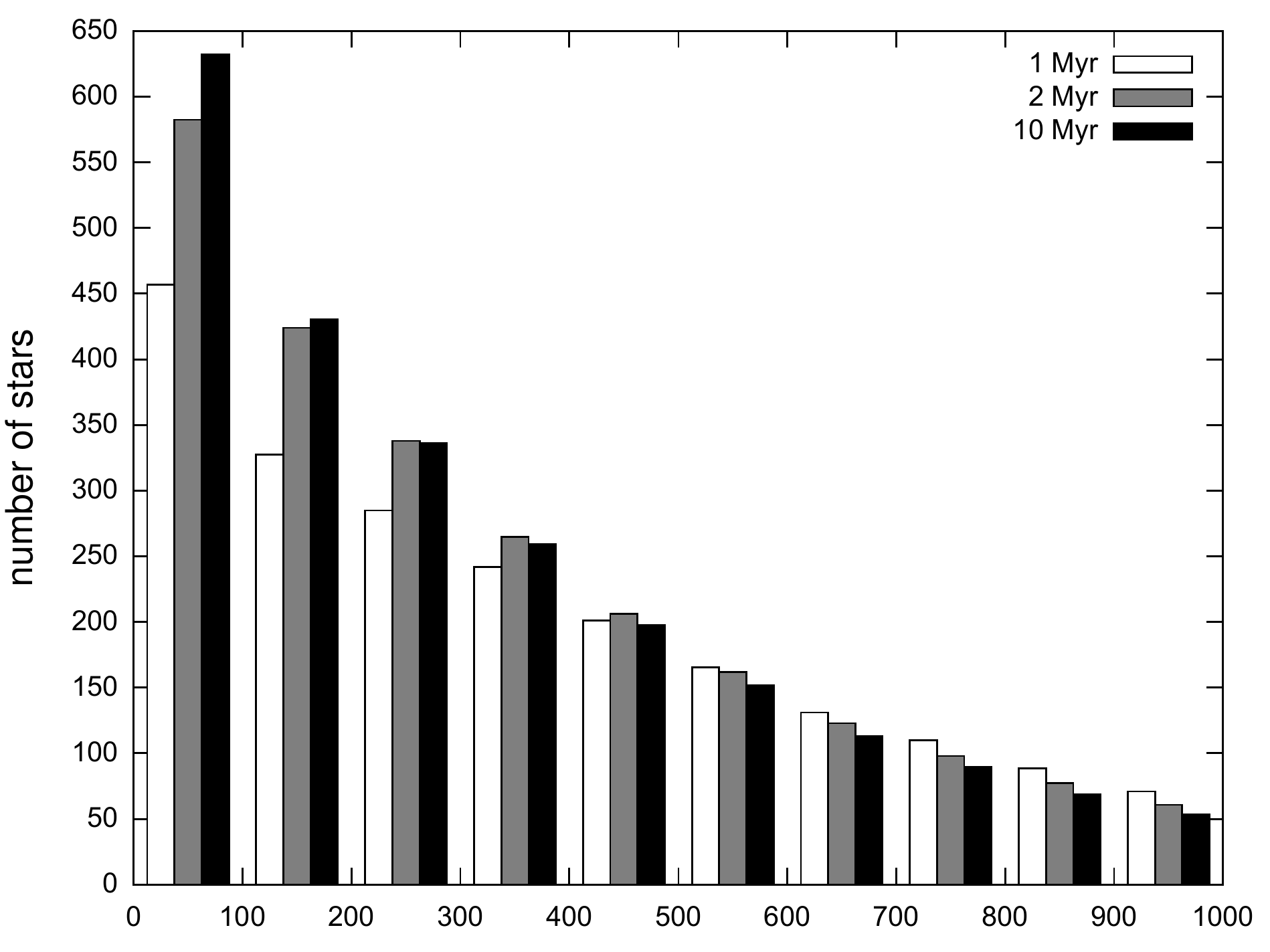}
  \caption{Disc-size distribution in the ONC-like cluster model (E2) for a fixed virtual FOV ($1~\text{pc}$) and different time steps: \mbox{$1~\text{Myr}$ (white)}, \mbox{$2~\text{Myr}$ (grey)}, and \mbox{$10~\text{Myr}$ (black)}.}
  \label{fig:disc_size_ONC_vs_time}
\end{figure}

  If observations of older clusters work like a microscope, what would an observed disc-size distribution in a cluster at different ages look like? 
  To answer this question, we investigate the ONC-model cluster (E2) at different ages with an artificial fixed Field of View (FOV) \mbox{($r_{FOV} = 1~\text{pc}$)} to mimic observations. Note that the FOV for observations are usually squares whereas here we present spheres with a radius of $r_{\text{FOV}}$, centred on the cluster origin.
  \mbox{Figure~\ref{fig:disc_size_ONC_vs_time}} shows the resulting disc-size statistics for an ONC-like cluster at $1~\text{Myr}$ (white), $2~\text{Myr}$ (grey), and $10~\text{Myr}$ (black).
  The total number of small discs increases much stronger than the number of discs with sizes of several hundreds of AU. The reason is that in the embedded phase discs which are already influenced but still a few hundreds of AU large still reduced in size by follow-up encounters. In comparison, the shape of the disc-size distribution barely changes between the end of the embedded phase and the end of the simulations.

  Observations usually study only the central areas of a cluster, because their the stellar density of cluster members is so high that member identification is relatively easy - basically the rate of false-positives is very low. However, this concentration on the cluster centre is problematic, especially so for clusters after gas expulsion, which span large areas.
  Taking our results as a guideline the observed median disc size in an ONC-like cluster $10~\text{Myr}$ after cluster formation, for example, would be \mbox{$\sim 50~\text{AU}$} in the cluster core $(0.2~\text{pc})$ whereas the overall median disc size is more than nine times as large ($\sim 460~\text{AU}$, dotted horizontal line in Figure~\ref{fig:mean_disc_size_vs_cluster}).

\begin{figure}[t]
  \centering
  \includegraphics[width=0.48\textwidth]{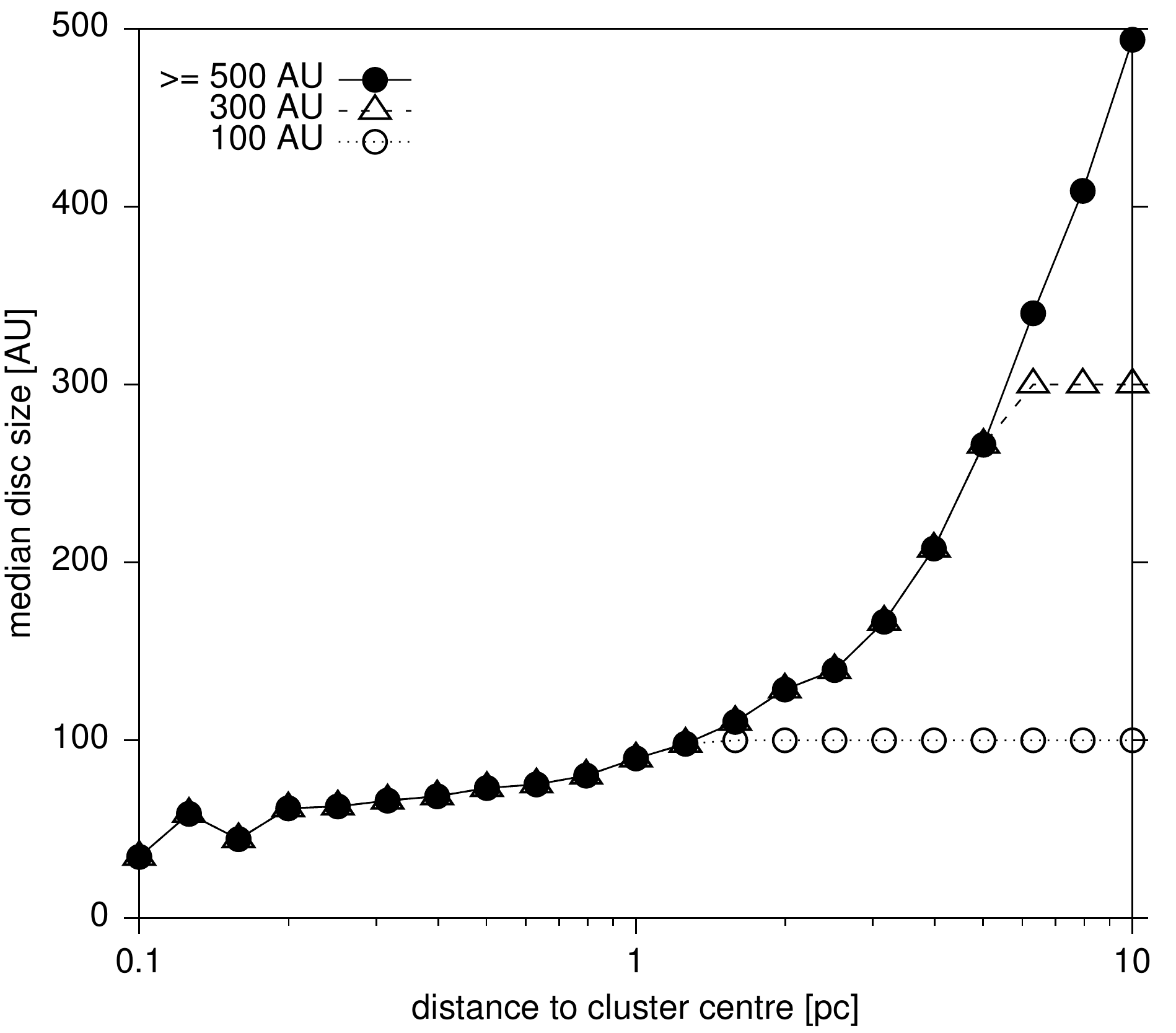}
  \caption{Median disc size at the end of the simulation ($10~\text{Myr}$) as a function of the distance to the cluster centre for cluster \mbox{model E2} for different initial disc sizes: \mbox{$100~\text{AU}$} (circles), \mbox{$300~\text{AU}$} (triangles), and \mbox{$\ge 500~\text{AU}$} (dots).}
  \label{fig:mean_disc_size_vs_distance_to_cluster_center_vs_initial_disc_size}
\end{figure}

  Choosing initially artificially large discs of $10\,000~\text{AU}$ has the advantage that the obtained results can be applied any smaller, real disc size. Thus, Figure~\ref{fig:mean_disc_size_vs_distance_to_cluster_center_vs_initial_disc_size}, tells us, for example, that if all stars had an initial disc size of \mbox{$r_{\text{init}} \geq 500~\text{AU}$}, about half the stars had their discs severely truncated by fly-bys to disc sizes below $500~\text{AU}$.
  An initial disc size of more than $500~\text{AU}$ is a realistic scenario as surveys found discs in the ONC with radii of $30-500~\text{AU}$ \citep{McCaughrean_Odell_1996, Vicente_Alves_2005, Eisner_et_al_2008, Bally_et_al_2015}. Note, that at an age of approximately $1~\text{Myr}$ even those might already have been reduced in size through photoevaporation and/or fly-bys.
  In the case of more massive clusters like NGC~6611 \mbox{(E52 model)} there are more and closer interactions, so that independent of the initial disc size (as long as $r_{\text{init}}>100~\text{AU}$) the resulting median disc is $\leq 110~\text{AU}$, see Fig.~\ref{fig:mean_disc_size_vs_cluster}. \\

  In summary, observed disc sizes or disc-size distributions in massive clusters are a strong function of the cluster age, its evolutionary stage, its initial conditions, and the FOV of the instrument. One has to act with caution when comparing and interpreting such results.

\section{Discussion}
\label{sec:discussion}

  The above described simulations required some approximations, which we discuss in the following. 
  
  In this study we neglect potentially existing initial substructuring of the clusters. In clusters with low velocity dispersions the substructure will be erased quickly \citep[see e.g.][]{Goodwin_Whitworth_2004, Allison_et_al_2010, Parker_et_al_2014}. Most probably, substructure will be erased at the end of star formation \citep{Bonnell_Bate_Vine_2003}, which is when our simulations start.
  
  The cluster models were set up without primordial mass segregation. Many clusters show signs of mass segregation, but it is unclear whether this property is primordial or if dynamical evolution caused the observed mass segregation. 
  If we included primordial mass segregation, the most massive stars would reside in the cluster core where the density is highest. Therefore, they would undergo more fly-bys, leading in turn to smaller discs around these stars. Furthermore, stronger gravitational focussing would lead to an increase in the overall fly-by frequency in the cluster centre and thus smaller discs.

  All stars in the clusters were set up to be initially single excluding primordial binaries. 
  Observations show that the multiplicity, that is the fraction of binaries, triples or systems of higher order, increases with stellar mass \citep[][and references therein]{Koehler_et_al_2006, Duchene_Kraus_2013}. 
  The most massive stars would most probably be part of a binary then, losing their own disc quite quickly or not even forming one depending on the separation.
  Additionally, the gravitational focussing in the cluster due to multiple systems would be stronger than for a single, massive star, leading to an increase in the fly-by frequency and overall smaller discs.
  
  One major difference to previous work is that we studied different evolutionary stages starting with the embedded phase, continuing with the gas expulsion, and the following expansion phase.
  Due to the uncertainty in the age determination of clusters, the duration of the embedded phase is not well constrained by observations. Here we modelled the duration of the embedded phase as $2~\text{Myr}$.
  However, for the most massive clusters this is probably too long, as at that age massive clusters are already largely devoid of gas. As most of the disc-size reducing fly-bys occur during the early stages of the embedded phase, with only $12\%$ of fly-bys happening in the second half of the embedded phase for the most massive clusters, our results should be not very sensitive to the assumed duration of the embedded phase.

  The assumption of instantaneous gas expulsion is most likely justified for the most massive clusters in our investigation \citep[e.g.][]{Geyer_Burkert_2001, Melioli_de_Gouveia_dal_Pino_2006, Portegies_Zwart_McMillan_Gieles_2010}. Nevertheless, for the lowest mass clusters this is less certain. In this case a “slow” gas expulsion lasting several million years would give the cluster more time to adjust to the gas-mass loss and fewer stars would become unbound. Furthermore, the stellar density would remain higher for a longer time span, allowing the stars to undergo more fly-bys and resulting in smaller discs than in the presented results. 
  The influence of only the embedded phase was studied in VBP15. Comparing this to our current work, we find that the duration of the embedded phase e.g. for the lowest density cluster\footnote{model D0 in VBP15, equivalent to model E0 here} is strong. At the end of the embedded phase in VBP15 - lasting unrealistically $5~\text{Myr}$ - the mean disc size is roughly $300~\text{AU}$ inside $0.6~\text{pc}$, compared to $\sim 670~\text{AU}$) for the here adopted $2~\text{Myr}$ long embedded phase. If the gas was expelled slowly, a more realistic median disc size would lie between these two extremes.
  Further studies with an implicitly modelled gas-expulsion of a few Myr are necessary to constrain this rough estimate. \\

  Here all fly-bys were assumed to be prograde, coplanar, and parabolic. Those fly-bys have a stronger effect on the discs than their retrograde, inclined counterparts \citep{Clarke_Pringle_1993, Heller_1995, Hall_1997, Pfalzner_et_al_2005, Bhandare_Pfalzner_2015}. However, \cite{Pfalzner_et_al_2005} found that for fly-bys with inclinations of $<45^{\circ}$ the final disc properties do not differ much from the prograde coplanar case. This was confirmed by \cite{Bhandare_Pfalzner_2015}, who found that even retrograde fly-bys can have a strong effect on the disc size. Discs after inclined fly-bys would be larger than the ones presented here, but at most by a factor of $1.5-1.8$.

  We only considered parabolic encounters, however, the typical eccentricity of a fly-by depends on the cluster density: the higher the density, the more eccentric are the fly-bys, see Fig.~\ref{app_fig:dsc_enc_vs_ecc_vs_time}. As pointed out by \cite{Pfalzner_2004} such hyperbolic fly-bys have less influence on discs than parabolic fly-bys, making the here presented disc sizes again lower limits. A detailed parameter study of hyperbolic fly-bys and their influence on the disc size would be necessary to extend our study.

  In this work we did not include photoevaporation, which is also capable of reducing discs in size or destroying them completely \citep[][]{Stoerzer_Hollenbach_1999, Scally_Clarke_2002, Johnstone_et_al_2004, Adams_et_al_2006, Alexander_Clarke_Pringle_2006, Ercolano_et_al_2008, Gorti_Hollenbach_2009, Drake_et_al_2009}. 
  In the embedded phase the stars are still surrounded by the cluster’s natal gas which makes the external photoevaporation ineffective. When the gas is expelled the discs are prone to the radiation from nearby massive stars.
  Nevertheless, the stars move outwards and may become unbound after the gas expulsion, the stellar density decreases significantly making it less probable for stars being very close to their most massive companions. Only for the small fraction of stars which have a close fly-by the radiation would further reduce the disc in size making the final discs sizes smaller than presented here.

  Here we study the effect on low-mass discs. In this case viscosity and self-gravity of the disc can be neglected during the encounter as such. However, viscosity would lead to disc spreading in the long-term.
  \cite{Rosotti_et_al_2014} performed combined Nbody/SPH-simulations of low-mass clusters including viscous discs and found viscous spreading a) counteracting the size reduction due to stellar fly-bys and b) making the discs prone to follow-up more distant fly-bys.
  Recently, \cite{Xiang-Gruess_2015} compared the results of Nbody and SPH simulations of discs after stellar fly-bys, showing that viscosity can result in warped disc structures whereas those features are not visible in mass-less (purely Nbody) discs. A disc-size determination in those cases would be more complicated than in the flat, mass-less discs used here.

  We did not consider any dependence of the disc size on the host's mass \citep[cf. e.g.][]{Hillenbrand_et_al_1998, Vicente_Alves_2005, Eisner_et_al_2008, Vorobyov_2011}. 
  If the initial disc size did depend on the stellar mass, the more massive stars should have started out with larger discs than the less massive stars. \cite{Vorobyov_2011} performed simulations of discs around \mbox{Class 0} and \mbox{Class I} stars. They set a density threshold of $\Sigma < 0.1 \text{g~cm}^{-2}$ for material belonging to the discs and found disc sizes between roughly $100~\text{AU}$ for low-mass stars up to a little more than $1\,000~\text{AU}$ for solar-like stars. If confirmed it would mean that discs around massive stars are more prone to size changes by the environment than low-mass stars.
  Furthermore, this would mean that in all clusters, except \mbox{model E0}, more than half of the discs around solar-like stars would be influenced strongly by stellar fly-bys, see \mbox{Fig.~\ref{fig:mean_disc_size_vs_cluster}}.
  
  Recent simulations have tried to determine the fraction of planets that become affected by the cluster environment and either move on an eccentric orbit or become unbound \citep{Hao_Kouwenhoven_Spurzem_2013,Li_Adams_2015}. However, these simulations concentrate on the initially much denser clusters that become long-lived open clusters. This type of cluster will be studied in a follow-up paper.

\section{Summary and Conclusion}
\label{sec:summary_and_conclusion}

  In this paper we studied how the cluster environment changes the sizes of discs surrounding young stars. In contrast to previous work we took the cluster development during the first $10~\text{Myr}$ explicitly into account. 
  Starting with initial conditions typical for young clusters at the end of their formation phase in the solar neighbourhood, we modelled the cluster dynamics from embedded throughout the expansion phase and determined the effect on the effect of gravitational interactions between the stars on the disc sizes. 
  These type of simulations were performed for clusters of different mass and density. \vspace{1em}

  \noindent{Our findings are the following:}
  \vspace{-0.5em}
  \begin{enumerate}
    \item It is essential to include the gas dynamics in this kind of simulations, as the larger velocity dispersion leads to more encounters and significantly smaller disc sizes than in a gas-free treatment.
    \item The majority of disc-size changing fly-bys always takes place in the embedded phase. However, the slower expansion phase in lower mass clusters means that here still $12\%$ of disc-size changing fly-bys happen, in comparison to just $2\%$ for high-mass clusters.
    \item For ONC-like clusters basically only discs larger than 500 AU are affected by fly-bys, whereas in NGC~6611-like clusters, cutting discs below 100 AU happens for $50\%$ of stars.
    \item However, in all investigated cases the disc sizes in the dense cluster centres are much more affected than the average suggests. For example, in the NGC~6611-like case the median disc size is 54 AU.
    \item The duration of the embedded phase influences the final median disc size, but not as strong as one would expect, because early fly-bys reduce the disc size already, leading to smaller cross sections for later fly-bys. In the densest cluster the median disc size after $1~\text{Myr}$ is already $155~\text{AU}$, at the end of the embedded phase $108~\text{AU}$, which is very close to the final median disc size of $104~\text{AU}$.
  \end{enumerate}

  Often disc sizes and frequencies \citep[e.g.][and references therein]{Haisch_Lada_Lada_2001, Mamajek_2009} of clusters of different present density are compared to obtain information about to what degree the environment influences these properties. However, clusters are highly dynamical and their current density is not necessarily representative for the past development. 
  We showed that between $3-4~\text{Myr}$ even the most extreme cluster models of E0 and E52 have very similar cluster mass densities within a sphere of $1.3~\text{pc}$\footnote{Note that the given simulation time is not synonymous with cluster age, as the star formation phase is not covered by out simulations}, see \mbox{Fig.~\ref{fig:cluster_mass_density_vs_time}}.
  The faster evolution of massive clusters leads this situation where the density in massive clusters and low-mass clusters of the same age can be similar, but the clusters themselves are in very different evolutionary stages.
  This means that at this specific point in time and in this sphere of $1.3~\text{pc}$ fly-bys are equally likely in all of these initially very different clusters.
  However, if we compare the median disc sizes in these ``equal-density'' clusters at this point in time ($t=4\text{Myr}$), they differ considerably. 
  In the least dense cluster the median disc size is roughly $480~\text{Myr}$ whereas it is $270~\text{AU}$ for the densest cluster model. The reason is that the most massive clusters where once much denser than the lower mass clusters and therefore their disc sizes are reduced to a larger degree. 
      
  The different expansion of the clusters - slow for low-density and fast for high-density systems - leads to very distinct fly-by histories and, consequently, different median disc sizes and disc-size distributions. If one looks only at the embedded phase there seems to be a direct relation between stellar density and the disc size: the higher the density, the smaller the median size.
  Thus it seems that this is easily testable against observations. However, taking into account the different evolutionary phases and their different time scales for dense and less dense clusters show that a comparison is much more complex.

  All these effects of cluster properties and observational constraints make it quite challenging to compare disc-size distributions in different clusters with each other. It does not make sense comparing the properties in clusters of different densities as long as one does not take into account their evolutionary stage and their history. 




\begin{appendix}
  \section{Fly-by velocity and eccentricity}
  \label{app:fly-by_eccentricity}
  
\begin{figure}[t]
  \centering
  \begin{subfigure}[t]{0.48\textwidth}
    \includegraphics[width=\textwidth]{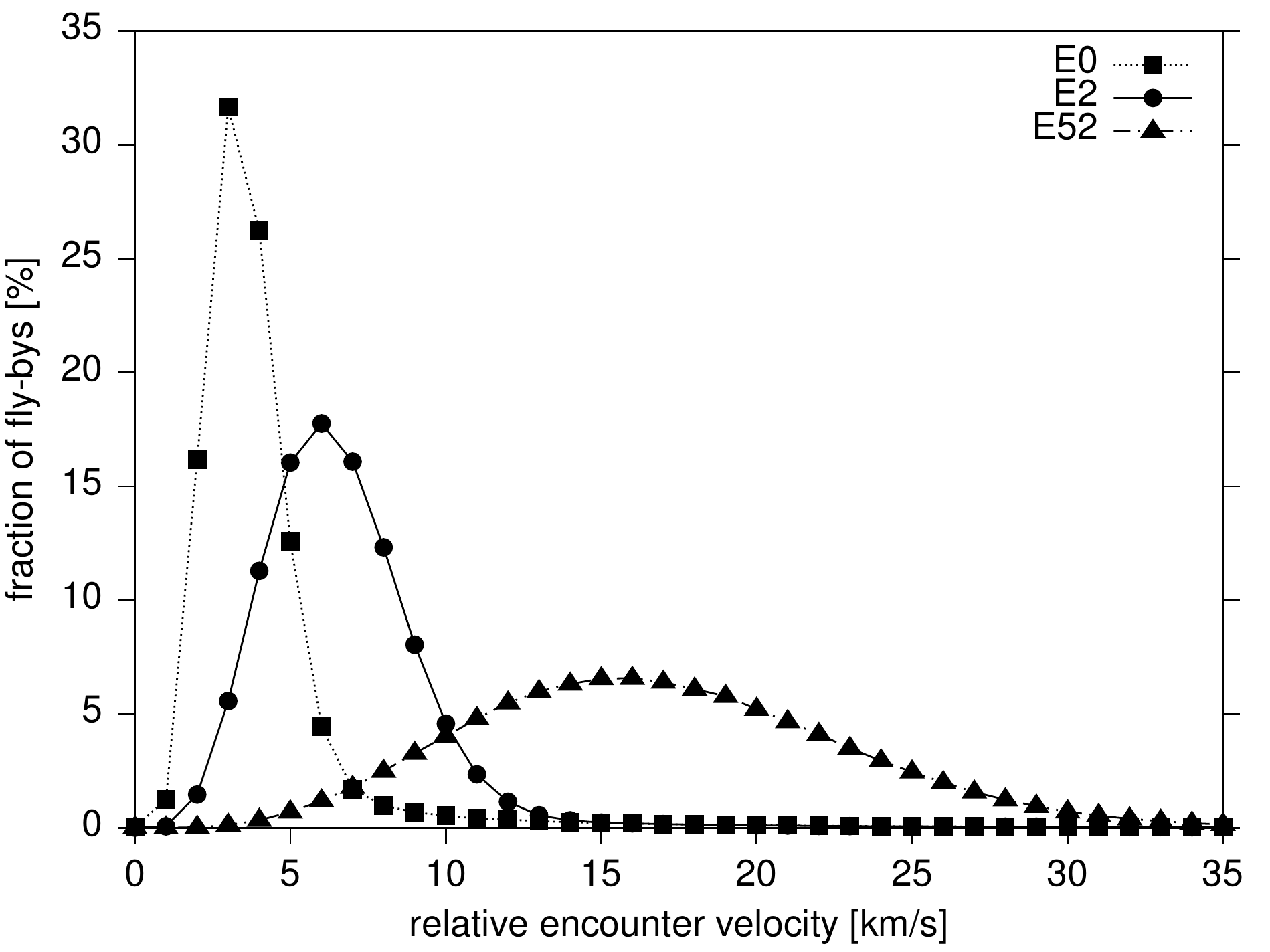}
  \end{subfigure}
  \hfill
  \begin{subfigure}[t]{0.48\textwidth}
    \includegraphics[width=\textwidth]{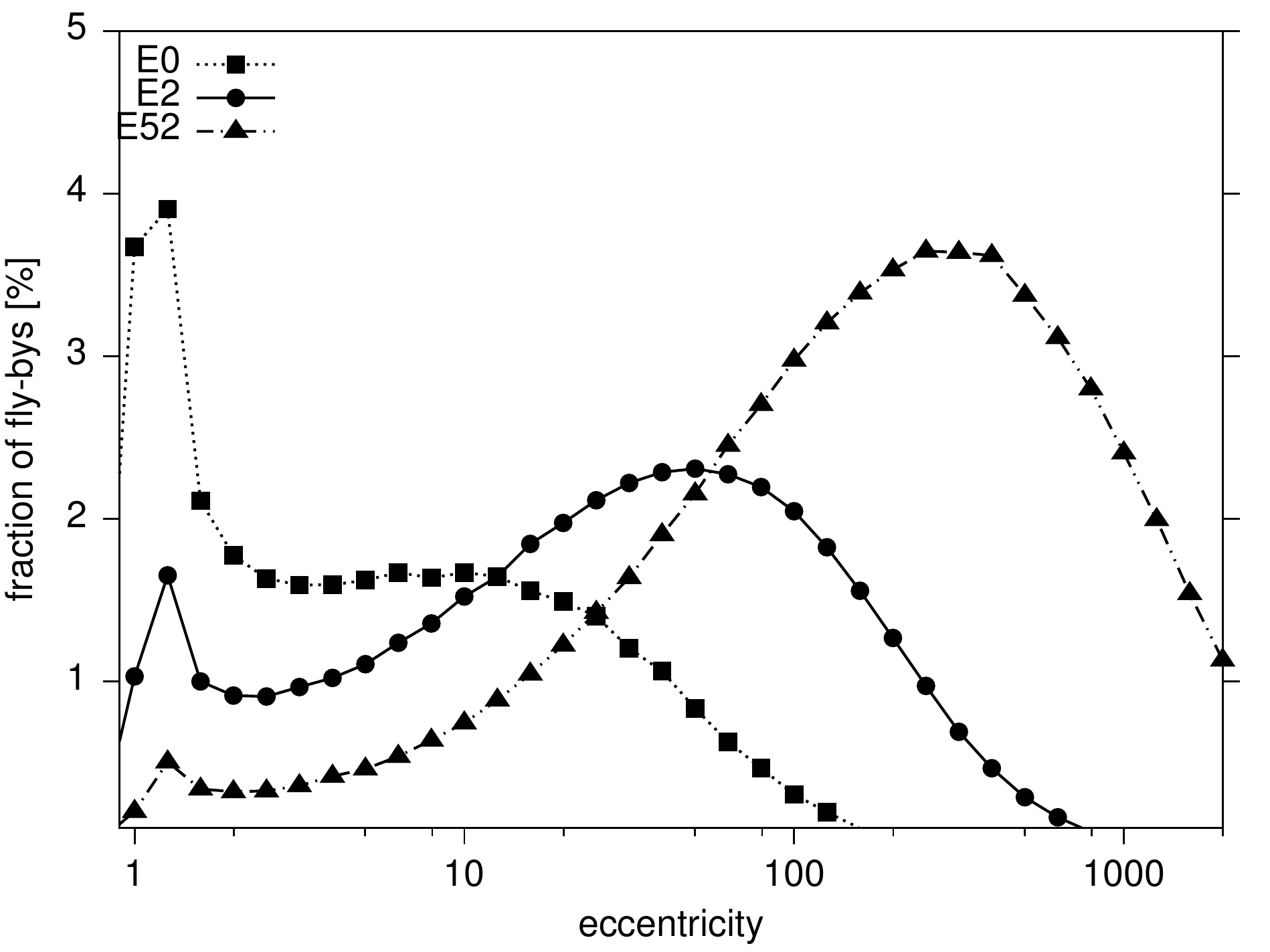}
  \end{subfigure}
  \caption{(a) relative encounter velocity distribution - that is the perturber's velocity relative the the host's velocity at the point of periastron passage - of all fly-bys and (b) eccentricity distribution of fly-bys leading to a disc size $<500~\text{AU}$ as fractions of the total number of fly-bys for three cluster models: \mbox{E0 (squares)}, \mbox{E2 (dots)}, and \mbox{E52 (triangles)}.}
  \label{app_fig:dsc_enc_vs_ecc_vs_time}
\end{figure}

    The characteristics of stellar encounters change significantly with cluster density. For example, the relative velocity between two encountering stars increases for denser clusters. Figure~\ref{app_fig:dsc_enc_vs_ecc_vs_time}(a) depicts the average relative encounter velocity - that is the velocity of the perturber relative to the host star at the time of periastron passage - for three cluster models \mbox{E0 (squares)}, \mbox{E2 (dots)}, and \mbox{E52 (triangles)}. This encounter velocity can directly be correlated to the eccentricity of the perturber's orbit via:
    
    The eccentricity $e$ is directly related to the relative encounter velocity of the two stars at the time of periastron passage:
    \begin{equation}
      v_{\text{enc}} = \sqrt{(1+e) \cdot \frac{G(m_1 + m_2)}{r_{\text{peri}}}},
      \label{eq:encounter_velocity}
    \end{equation}
    
    where $G$ is the gravitational constant, $m_1$ is the mass of the host star, $m_2$ the mass of the perturber, and $r_{\text{peri}}$ the periastron distance, all in SI units. The eccentricity distribution for cluster models E0, E2, and E52 are shown in Fig.~\ref{app_fig:dsc_enc_vs_ecc_vs_time}(b) for fly-bys leading to disc smaller than $500~\text{AU}$.
    
    In this study we assumed all fly-bys to be parabolic. 
    This approximation only holds for the least dense cluster model, as the encounter velocities and therefore the eccentricities clearly increases with cluster density \citep[see also][]{Olczak_Pfalzner_Eckart_2010}. For the denser cluster models (especially E52) a detailed study of the influence of hyperbolic fly-bys on disc sizes would be favourable. 
    Previous studies suggest that their influence on the discs (in these cases the disc mass and angular momentum) is much smaller than the one of parabolic encounters \citep[for detailed discussions see e.g.][]{Pfalzner_et_al_2005, Olczak_Pfalzner_Eckart_2010, Olczak_et_al_2012}. Therefore, the disc sizes presented here might be lower limits.
    
    At very high densities, that is especially in cluster model E52, fly-bys are no longer 2-body encounters but many-body interactions. This leads to the extreme eccentricities of $e>100$. Especially for this type of fly-by we expect the disc-size change to be smaller than for the here assumed prograde, coplanar, parabolic case.

\end{appendix}

\clearpage

\clearpage

\begin{table}
  \begin{center}
    \caption{Cluster model set-up and dynamical time scales.}
    \begin{tabular}[t]{crrrrrrr} \tableline \tableline
      Model  & $N_{\text{stars}}$  & $N_{\text{sim}}$  & $t_{\text{emb}}$  & $r_{\text{hm}}$  & $M_{\text{stars}}$  & $M_{\text{cl}}$   & $t_{\text{dyn}}$  \\
             &                     &                   & [Myr]             & [pc]             & $[M_{\sun}]$        & $[M_{\sun}]$      & [Myr]             \\ \tableline
      E0     & $1\,000$            & 308               & 2.0               & 1.3              & $590.8$             & $1\,969.2$        & 0.67              \\ 
      E1     & $2\,000$            & 168               & 2.0               & 1.3              & $1\,192.2$          & $3\,973.9$        & 0.47              \\ 
      E2     & $4\,000$            & 94                & 2.0               & 1.3              & $2\,358.1$          & $7\,860.3$        & 0.33              \\ 
      E3     & $8\,000$            & 47                & 2.0               & 1.3              & $4\,731.2$          & $15\,770.6$       & 0.24              \\ 
      E4     & $16\,000$           & 16                & 2.0               & 1.3              & $9\,464.8$          & $31\,549.3$       & 0.17              \\ 
      E52    & $32\,000$           & 9                 & 2.0               & 1.3              & $18\,852.6$         & $62\,842.0$       & 0.12              \\
      E51    & $32\,000$           & 7                 & 1.0               & 1.3              & $18\,839.2$         & $62\,797.3$       & 0.12              \\
    \tableline
    \end{tabular}
    \tablecomments{Column~1 indicates the model designation, followed by the initial number of stars in the cluster $N_{\text{stars}}$, the number of simulations in campaign $N_{\text{sim}}$, the duration of the embedded phase $t_{\text{emb}}$, the initial half-mass radius $r_{\text{hm}}$ of the cluster, the stellar mass of the cluster $M_{\text{stars}}$, the total cluster mass (including the gas mass) $M_{\text{cl}}$, and the resulting dynamical time scale $t_{\text{dyn}}$. For calculation of $M_{\text{cl}}$ and $t_{\text{dyn}}$, see text.}
     \label{tab:set-up_params}
  \end{center}
\end{table}

\clearpage



\begin{thebibliography}{}
  \expandafter\ifx\csname natexlab\endcsname\relax\def\natexlab#1{#1}\fi

  \bibitem[{{Aarseth}(1973)}]{Aarseth_1973}{Aarseth}, S.~J. 1973, Vistas in Astronomy, 15, 13

  \bibitem[{{Aarseth}(2003)}]{Aarseth_2003}{Aarseth}, S.~J. 2003, {Gravitational N-Body Simulations}

  \bibitem[{{Adams}(2000)}]{Adams_2000}{Adams}, F.~C. 2000, \apj, 542, 964

  \bibitem[{{Adams}(2010)}]{Adams_2010}{Adams}, F.~C. 2010, \araa, 48, 47

  \bibitem[{{Adams} {et~al.}(2006){Adams}, {Proszkow}, {Fatuzzo}, \& {Myers}}]{Adams_et_al_2006}{Adams}, F.~C., {Proszkow}, E.~M., {Fatuzzo}, M., \& {Myers}, P.~C. 2006, \apj, 641, 504

  \bibitem[{{Alexander} {et~al.}(2006){Alexander}, {Clarke}, \& {Pringle}}]{Alexander_Clarke_Pringle_2006}{Alexander}, R.~D., {Clarke}, C.~J., \& {Pringle}, J.~E. 2006, \mnras, 369, 229
  
  \bibitem[{{Allison} {et~al.}(2010){Allison}, {Goodwin}, {Parker}, {Portegies Zwart} \& {de Grijs}}]{Allison_et_al_2010}{Allison}, R.~J., {Goodwin}, S.~P., {Parker}, R.~J., {Portegies Zwart}, S.~F., \& {de Grijs} 2010, \mnras, 407, 1098

  \bibitem[{{Balbus} \& {Hawley}(2002)}]{Balbus_Hawley_2002}{Balbus}, S.~A. \& {Hawley}, J.~F. 2002, \apj, 573, 749

  \bibitem[{{Bally} {et~al.}(2015){Bally}, {Mann}, {Eisner}, {Andrews}, {Di Francesco}, {Hughes}, {Johnstone}, {Matthews}, {Ricci}, \&  {Williams}}]{Bally_et_al_2015}{Bally}, J., {Mann}, R.~K., {Eisner}, J., {et~al.} 2015, ArXiv e-prints

  \bibitem[{{Baumgardt} \& {Kroupa}(2007)}]{Baumgardt_Kroupa_2007}{Baumgardt}, H. \& {Kroupa}, P. 2007, \mnras, 380, 1589

  \bibitem[{{Bhandare} \& {Pfalzner}({subm.})}]{Bhandare_Pfalzner_2015}{Bhandare}, A. \& {Pfalzner}, S. {in prep.}

  \bibitem[{{Boily} \& {Kroupa}(2003{\natexlab{a}})}]{Boily_Kroupa_2003a}{Boily}, C.~M. \& {Kroupa}, P. 2003{\natexlab{a}}, \mnras, 338, 665

  \bibitem[{{Boily} \& {Kroupa}(2003{\natexlab{b}})}]{Boily_Kroupa_2003b}{Boily}, C.~M. \& {Kroupa}, P. 2003{\natexlab{b}}, \mnras, 338, 673

  \bibitem[{{Bonnell} {et~al.}(2003){Bonnell}, {Bate}, \& {Vine}}]{Bonnell_Bate_Vine_2003}{Bonnell}, I.~A., {Bate}, M.~R., {Vine}, S.~G. 2003, \mnras, 343, 413
  
  \bibitem[{{Bonnell} \& {Davies}(1998)}]{Bonnell_Davies_1998}{Bonnell}, I.~A. \& {Davies}, M.~B. 1998, \mnras, 295, 691

  \bibitem[{{Breslau} {et~al.}(2014){Breslau}, {Steinhausen}, {Vincke}, \&  {Pfalzner}}]{Breslau_et_al_2014}{Breslau}, A., {Steinhausen}, M., {Vincke}, K., \& {Pfalzner}, S. 2014, \aap,  565, A130

  \bibitem[{{Clarke} \& {Pringle}(1993)}]{Clarke_Pringle_1993}{Clarke}, C.~J. \& {Pringle}, J.~E. 1993, \mnras, 261, 190

  \bibitem[{{Dale} {et~al.}(2012){Dale}, {Ercolano}, \&  {Bonnell}}]{Dale_Ercolano_Bonnell_2012}{Dale}, J.~E., {Ercolano}, B., \& {Bonnell}, I.~A. 2012, \mnras, 424, 377

  \bibitem[{{Dale} {et~al.}(2015){Dale}, {Ercolano}, \&  {Bonnell}}]{Dale_Ercolano_Bonnell_2015}{Dale}, J.~E., {Ercolano}, B., \& {Bonnell}, I.~A. 2015, \mnras, 451, 5506

  \bibitem[{{de Juan Ovelar} {et~al.}(2012){de Juan Ovelar}, {Kruijssen},{Bressert}, {Testi}, {Bastian}, \& {C{\'a}novas}}]{de_Juan_Ovelar_et_al_2012}{de Juan Ovelar}, M., {Kruijssen}, J.~M.~D., {Bressert}, E., {et~al.} 2012,  \aap, 546, L1

  \bibitem[{{Drake} {et~al.}(2009){Drake}, {Ercolano}, {Flaccomio}, \&  {Micela}}]{Drake_et_al_2009}{Drake}, J.~J., {Ercolano}, B., {Flaccomio}, E., \& {Micela}, G. 2009, \apjl,  699, L35

  \bibitem[{{Duch{\^e}ne} \& {Kraus}(2013)}]{Duchene_Kraus_2013}{Duch{\^e}ne}, G. \& {Kraus}, A. 2013, \araa, 51, 269

  \bibitem[Dukes \& Krumholz(2012)]{Dukes_Krumholz_2012} Dukes, D., \& Krumholz, M.~R.\ 2012, \apj, 754, 56

  \bibitem[{{Eisner} {et~al.}(2008){Eisner}, {Plambeck}, {Carpenter}, {Corder},  {Qi}, \& {Wilner}}]{Eisner_et_al_2008}{Eisner}, J.~A., {Plambeck}, R.~L., {Carpenter}, J.~M., {et~al.} 2008, \apj,
  683, 304

  \bibitem[{{Ercolano} {et~al.}(2008){Ercolano}, {Drake}, {Raymond}, \&  {Clarke}}]{Ercolano_et_al_2008}{Ercolano}, B., {Drake}, J.~J., {Raymond}, J.~C., \& {Clarke}, C.~C. 2008,  \apj, 688, 398

  \bibitem[Fall et al.(2009)]{Fall_Chandar_Whitmore_2009} Fall, S.~M., Chandar, R., \& Whitmore, B.~C.\ 2009, \apj, 704, 453

  \bibitem[{{Fellhauer} \& {Kroupa}(2005)}]{Fellhauer_Kroupa_2005}{Fellhauer}, M. \& {Kroupa}, P. 2005, \apj, 630, 879

  \bibitem[{{Garc{\'{\i}}a} {et~al.}(2014){Garc{\'{\i}}a}, {Bronfman}, {Nyman}, {Dame}, \& {Luna}}]{Garcia_et_al_2014}{Garc{\'{\i}}a}, P. and {Bronfman}, L. and {Nyman}, L.-A. {et~al.} 2014, \apjs, 212, 2 
 
  \bibitem[{{Geyer} \& {Burkert}(2001)}]{Geyer_Burkert_2001}{Geyer}, M.~P. \& {Burkert}, A. 2001, \mnras, 323, 988

  \bibitem[{{Goodwin}(1997)}]{Goodwin_1997}{Goodwin}, S.~P. 1997, \mnras, 284, 785

  \bibitem[{{Goodwin} \& {Bastian}(2006)}]{Goodwin_Bastian_2006}{Goodwin}, S.~P. \& {Bastian}, N. 2006, \mnras, 373, 752

  \bibitem[{{Goodwin} \& {Whitworth}(2004)}]{Goodwin_Whitworth_2004}{Goodwin}, S.~P. \& {Whitworth}, A.~P. 2004, \aap 413, 929

  \bibitem[{{Gorti} \& {Hollenbach}(2009)}]{Gorti_Hollenbach_2009}{Gorti}, U. \& {Hollenbach}, D. 2009, \apj, 690, 1539

  \bibitem[{{Haisch} {et~al.}(2001){Haisch}, {Lada}, \&  {Lada}}]{Haisch_Lada_Lada_2001}{Haisch}, Jr., K.~E., {Lada}, E.~A., \& {Lada}, C.~J. 2001, \apjl, 553, L153

  \bibitem[{{Hall}(1997)}]{Hall_1997}{Hall}, S.~M. 1997, \mnras, 287, 148

  \bibitem[Hao et al.(2013)]{Hao_Kouwenhoven_Spurzem_2013} Hao, W., Kouwenhoven, M.~B.~N., \& Spurzem, R.\ 2013, \mnras, 433, 867

  \bibitem[{{Heller}(1995)}]{Heller_1995}{Heller}, C.~H. 1995, \apj, 455, 252

  \bibitem[{{Hillenbrand} {et~al.}(1998){Hillenbrand}, {Strom}, {Calvet},  {Merrill}, {Gatley}, {Makidon}, {Meyer}, \&  {Skrutskie}}]{Hillenbrand_et_al_1998}{Hillenbrand}, L.~A., {Strom}, S.~E., {Calvet}, N., {et~al.} 1998, \aj, 116,  1816

  \bibitem[{{Huff} \& {Stahler}(2006)}]{Huff_Stahler_2006}{Huff}, E.~M. \& {Stahler}, S.~W. 2006, \apj, 644, 355

  \bibitem[{{Johnstone} {et~al.}(1998){Johnstone}, {Hollenbach}, \&  {Bally}}]{Johnstone_Hollenbach_Bally_1998}{Johnstone}, D., {Hollenbach}, D., \& {Bally}, J. 1998, \apj, 499, 758

  \bibitem[{{Johnstone} {et~al.}(2004){Johnstone}, {Matsuyama}, {McCarthy}, \&  {Font}}]{Johnstone_et_al_2004}{Johnstone}, D., {Matsuyama}, I., {McCarthy}, I.~G., \& {Font}, A.~S. 2004, in  Revista Mexicana de Astronomia y Astrofisica Conference Series, Vol.~22,  Revista Mexicana de Astronomia y Astrofisica Conference Series, ed.  G.~{Garcia-Segura}, G.~{Tenorio-Tagle}, J.~{Franco}, \& H.~W. {Yorke}, 38--41

  \bibitem[{{Klahr} \& {Bodenheimer}(2003)}]{Klahr_Bodenheimer_2003}{Klahr}, H.~H. \& {Bodenheimer}, P. 2003, \apj, 582, 869

  \bibitem[{{Kobayashi} \& {Ida}(2001)}]{Kobayashi_Ida_2001}{Kobayashi}, H. \& {Ida}, S. 2001, \icarus, 153, 416

  \bibitem[{{K{\"o}hler} {et~al.}(2006){K{\"o}hler}, {Petr-Gotzens},  {McCaughrean}, {Bouvier}, {Duch{\^e}ne}, {Quirrenbach}, \&  {Zinnecker}}]{Koehler_et_al_2006}{K{\"o}hler}, R., {Petr-Gotzens}, M.~G., {McCaughrean}, M.~J., {et~al.} 2006,  \aap, 458, 461

  \bibitem[{{Kroupa}(2002)}]{Kroupa_2002}{Kroupa}, P. 2002, Science, 295, 82

  \bibitem[{{Kroupa}(2005)}]{Kroupa_2005}{Kroupa}, P. 2005, in ESA Special Publication, Vol. 576, The Three-Dimensional  Universe with Gaia, ed. C.~{Turon}, K.~S. {O'Flaherty}, \& M.~A.~C.  {Perryman}, 629

  \bibitem[{{Kroupa} {et~al.}(2001){Kroupa}, {Aarseth}, \&  {Hurley}}]{Kroupa_Aarseth_Hurley_2001}{Kroupa}, P., {Aarseth}, S., \& {Hurley}, J. 2001, \mnras, 321, 699

  \bibitem[{{Lada} \& {Lada}(2003)}]{Lada_Lada_2003}{Lada}, C.~J. \& {Lada}, E.~A. 2003, \araa, 41, 57

  \bibitem[{{Lada} {et~al.}(1984){Lada}, {Margulis}, \&  {Dearborn}}]{Lada_Margulis_Dearborn_1984}{Lada}, C.~J., {Margulis}, M., \& {Dearborn}, D. 1984, \apj, 285, 141

  \bibitem[{{Leisawitz} {et~al.}(1989){Leisawitz}, {Bash}, \&  {Thaddeus}}]{Leisawitz_Bash_Thaddeus_1989}{Leisawitz}, D., {Bash}, F.~N., \& {Thaddeus}, P. 1989, \apjs, 70, 731

  \bibitem[Li \& Adams(2015)]{Li_Adams_2015} Li, G., \& Adams, F.~C.\ 2015, \mnras, 448, 344

  \bibitem[{{L{\"u}ghausen} {et~al.}(2012){L{\"u}ghausen}, {Parmentier},  {Pflamm-Altenburg}, \& {Kroupa}}]{Lueghausen_et_at_2012}{L{\"u}ghausen}, F., {Parmentier}, G., {Pflamm-Altenburg}, J., \& {Kroupa}, P.  2012, \mnras, 423, 1985

  \bibitem[{{Mamajek}(2009)}]{Mamajek_2009}{Mamajek}, E.~E. 2009, in American Institute of Physics Conference Series, Vol.  1158, American Institute of Physics Conference Series, ed. T.~{Usuda},  M.~{Tamura}, \& M.~{Ishii}, 3--10

  \bibitem[{{Matsuyama} {et~al.}(2003){Matsuyama}, {Johnstone}, \&  {Hartmann}}]{Matsuyama_Johnstone_Hartmann_2003}{Matsuyama}, I., {Johnstone}, D., \& {Hartmann}, L. 2003, \apj, 582, 893

  \bibitem[{{Matzner} \& {McKee}(2000)}]{Matzner_McKee_2000}{Matzner}, C.~D. \& {McKee}, C.~F. 2000, \apj, 545, 364

  \bibitem[{{McCaughrean} \& {O'dell}(1996)}]{McCaughrean_Odell_1996}{McCaughrean}, M.~J. \& {O'dell}, C.~R. 1996, \aj, 111, 1977

  \bibitem[{{Melioli} \& {de Gouveia dal  Pino}(2006)}]{Melioli_de_Gouveia_dal_Pino_2006}{Melioli}, C. \& {de Gouveia dal Pino}, E.~M. 2006, \aap, 445, L23

  \bibitem[{{Moeckel} {et~al.}(2012){Moeckel}, {Holland}, {Clarke}, \&  {Bonnell}}]{Moeckel_et_al_2012}{Moeckel}, N., {Holland}, C., {Clarke}, C.~J., \& {Bonnell}, I.~A. 2012,  \mnras, 425, 450

  \bibitem[{{Murray}(2011)}]{Murray_2011}{Murray}, N. 2011, \apj, 729, 133
  
  \bibitem[{{Olczak} {et~al.}(2010){Olczak}, {Pfalzner}, \&  {Eckart}}]{Olczak_Pfalzner_Eckart_2010}{Olczak}, C., {Pfalzner}, S., \& {Eckart}, A. 2010, \aap, 509, A63

  \bibitem[{{Olczak} {et~al.}(2006){Olczak}, {Pfalzner}, \&  {Spurzem}}]{Olczak_Pfalzner_Spurzem_2006}{Olczak}, C., {Pfalzner}, S., \& {Spurzem}, R. 2006, \apj, 642, 1140

  \bibitem[{{Olczak} {et~al.}(2011){Olczak}, {Spurzem}, \&  {Henning}}]{Olczak_Spurzem_Henning_2011}{Olczak}, C., {Spurzem}, R., \& {Henning}, T. 2011, \aap, 532, A119

  \bibitem[{{Olczak} {et~al.}(2012){Olczak}, {Kaczmarek}, {Harfst}, {Pfalzner} \&  {Portegies Zwart}}]{Olczak_et_al_2012}{Olczak}, C., {Kaczmarek}, T., {Harfst}, S., {Pfalzner}, S., \& {Portegies Zwart}, S. 2012, \apj, 756, 123

  \bibitem[{{Parker} {et~al.}(2014){Parker}, {Wright}, {Goodwin} \&  {Meyer}}]{Parker_et_al_2014}{Parker}, R.~J., {Wright}, N.~J., {Goodwin}, S.~P. \& {Meyer}, M.~R. 2014, \mnras, 438, 620

  \bibitem[{{Parmentier} \& {Baumgardt}(2012)}]{Parmentier_Baumgardt_2012}{Parmentier}, G. \& {Baumgardt}, H. 2012, \mnras, 427, 1940
  
  \bibitem[{{Pelupessy} \& {Portegies  Zwart}(2012)}]{Pelupessy_Portegies_Zwart_2012}{Pelupessy}, F.~I. \& {Portegies Zwart}, S. 2012, \mnras, 420, 1503

  \bibitem[{{Pfalzner}(2004)}]{Pfalzner_2004}{Pfalzner}, S. 2004, \apj, 602, 356

  \bibitem[{{Pfalzner} \&  {Kaczmarek}(2013{\natexlab{a}})}]{Pfalzner_Kaczmarek_2013a}{Pfalzner}, S. \& {Kaczmarek}, T. 2013{\natexlab{a}}, \aap, 555, A135

  \bibitem[{{Pfalzner} \&  {Kaczmarek}(2013{\natexlab{b}})}]{Pfalzner_Kaczmarek_2013b}{Pfalzner}, S. \& {Kaczmarek}, T. 2013{\natexlab{b}}, \aap, 559, A38

  \bibitem[{{Pfalzner} \& {Olczak}(2007)}]{Pfalzner_Olczak_2007}{Pfalzner}, S. \& {Olczak}, C. 2007, \aap, 462, 193

  \bibitem[{{Pfalzner} {et~al.}(2006){Pfalzner}, {Olczak}, \&  {Eckart}}]{Pfalzner_Olczak_Eckart_2006}{Pfalzner}, S., {Olczak}, C., \& {Eckart}, A. 2006, \aap, 454, 811

  \bibitem[{{Pfalzner} {et~al.}(2014){Pfalzner}, {Parmentier}, {Steinhausen}, {Vincke}, \& {Menten}}]{Pfalzner_et_al_2014}{Pfalzner}, S., {Parmentier}, G., {Steinhausen}, M., {Vincke}, K. \& {Menten}, K. 2014, \apj, 794, 147
  
  \bibitem[{{Pfalzner} {et~al.}(2014){Pfalzner}, {Steinhausen}, \&  {Menten}}]{Pfalzner_Steinhausen_Menten_2014}{Pfalzner}, S., {Steinhausen}, M., \& {Menten}, K. 2014, \apjl, 793, L34

  \bibitem[{{Pfalzner} {et~al.}(2005{\natexlab{a}}){Pfalzner}, {Umbreit}, \&  {Henning}}]{Pfalzner_Umbreit_Henning_2005}{Pfalzner}, S., {Umbreit}, S., \& {Henning}, T. 2005{\natexlab{a}}, \apj, 629,  526

  \bibitem[{{Pfalzner} {et~al.}(2015){Pfalzner}, {Vincke}, \&  {Xiang}}]{Pfalzner_Vincke_Xiang_2015}{Pfalzner}, S., {Vincke}, K., \& {Xiang}, M. 2015, ArXiv e-prints

  \bibitem[{{Pfalzner} {et~al.}(2005{\natexlab{b}}){Pfalzner}, {Vogel},  {Scharw{\"a}chter}, \& {Olczak}}]{Pfalzner_et_al_2005}{Pfalzner}, S., {Vogel}, P., {Scharw{\"a}chter}, J., \& {Olczak}, C. 2005{\natexlab{b}}, \aap, 437, 967

 \bibitem[{Portegies Zwart}(2016)]{Portegies_Zwart_2016} Portegies Zwart, S.~F.\ 2016, \mnras, 457, 313

  \bibitem[{{Portegies Zwart} {et~al.}(2010){Portegies Zwart}, {McMillan}, \&  {Gieles}}]{Portegies_Zwart_McMillan_Gieles_2010}{Portegies Zwart}, S.~F., {McMillan}, S.~L.~W., \& {Gieles}, M. 2010, \araa,
  48, 431

  \bibitem[{{Rosotti} {et~al.}(2014){Rosotti}, {Dale}, {de Juan Ovelar},  {Hubber}, {Kruijssen}, {Ercolano}, \& {Walch}}]{Rosotti_et_al_2014}{Rosotti}, G.~P., {Dale}, J.~E., {de Juan Ovelar}, M., {et~al.} 2014, \mnras,  441, 2094

  \bibitem[{{Scally} \& {Clarke}(2001)}]{Scally_Clarke_2001}{Scally}, A. \& {Clarke}, C. 2001, \mnras, 325, 449

  \bibitem[{{Scally} \& {Clarke}(2002)}]{Scally_Clarke_2002}{Scally}, A. \& {Clarke}, C. 2002, \mnras, 334, 156

  \bibitem[Shara et al.(2016)]{2016ApJ...816...59S} Shara, M.~M., Hurley, J.~R., \& Mardling, R.~A.\ 2016, \apj, 816, 59

  \bibitem[{{Shu} {et~al.}(1987){Shu}, {Adams}, \&  {Lizano}}]{Shu_Adams_Lizano_1987}{Shu}, F.~H., {Adams}, F.~C., \& {Lizano}, S. 1987, \araa, 25, 23

  \bibitem[{{Spurzem}(1999)}]{Spurzem_1999}{Spurzem}, R. 1999, Journal of Computational and Applied Mathematics, 109, 407

  \bibitem[{{Steinhausen}(2013)}]{Steinhausen_phd_2013}{Steinhausen}, M. 2013, PhD thesis, Mathematisch-Naturwissenschaftliche  Fakult\"at der Universit\"at zu K\"oln

  \bibitem[{{Steinhausen} \& {Pfalzner}(2014)}]{Steinhausen_Pfalzner_2014}{Steinhausen}, M. \& {Pfalzner}, S. 2014, \aap, 565, A32

  \bibitem[{{St{\"o}rzer} \& {Hollenbach}(1999)}]{Stoerzer_Hollenbach_1999}{St{\"o}rzer}, H. \& {Hollenbach}, D. 1999, \apj, 515, 669

  \bibitem[{{Vicente} \& {Alves}(2005)}]{Vicente_Alves_2005}{Vicente}, S.~M. \& {Alves}, J. 2005, \aap, 441, 195

  \bibitem[{{Vincke} {et~al.}(2015){Vincke}, {Breslau}, \&  {Pfalzner}}]{Vincke_Breslau_Pfalzner_2015}{Vincke}, K., {Breslau}, A., \& {Pfalzner}, S. 2015, \aap

  \bibitem[{{Vorobyov}(2011)}]{Vorobyov_2011}{Vorobyov}, E.~I. 2011, \apj, 729, 146

  \bibitem[{{Xiang-Gruess}(2015)}]{Xiang-Gruess_2015}{Xiang-Gruess}, M. 2015, ArXiv e-prints

  \bibitem[{{Zwicky}(1953)}]{Zwicky_1953}{Zwicky}, F. 1953, \pasp, 65, 205

\end{thebibliography}
\end{document}